%% file: main.tex
\documentclass{article}

\usepackage[preprint]{neurips_2026}

\usepackage[utf8]{inputenc} 
\usepackage[T1]{fontenc}    
\usepackage{amsmath}        
\usepackage{hyperref}       
\usepackage{url}            
\usepackage{booktabs}       
\usepackage{amsfonts}       
\usepackage{nicefrac}       
\usepackage{microtype}      
\usepackage{xcolor}         
\usepackage{graphicx}       

\input{header}

\title{CUCo: An Agentic Framework for Compute and Communication Co-design}

\author{%
  Yoga Sri Varshan Varadharajan\thanks{Both authors contributed equally to this research.} \\
  UT Austin \\
  \texttt{yogasrivarshan@utexas.edu} \\
  \And
  Bodun Hu\footnotemark[1] \\
  UT Austin \\
  \texttt{bodunhu@utexas.edu} \\
  \And
  Saurabh Agarwal \\
  UT Austin \\
  \texttt{saurabh.agarwal@utexas.edu} \\
  \And
  Aditya Akella \\
  UT Austin \\
  \texttt{aditya@cs.utexas.edu} \\
}

\begin{document}

\maketitle

\begin{abstract}
Computation and communication in distributed LLM training and inference are traditionally optimized in isolation. Expert-crafted systems such as DeepEP, FLUX, and TokenWeave have demonstrated the potential of co-design, but each demands deep systems expertise and is tuned for specific hardware and workload configurations. We present \cufuse{}, an agentic framework that automates compute-communication co-design of CUDA kernels. \cufuse{} combines a structured design-space formalization with a correctness-first fast-path agent that produces reliable baselines and an evolution-driven slow-path agent that discovers high-performance co-design strategies. Across four multi-GPU workloads, \cufuse{} achieves up to 1.57$\times$ speedup over host-driven baselines. On a DeepSeek-V3 MoE layer, \cufuse{} discovers a two-stream overlap strategy that outperforms sequential DeepEP by hiding dispatch behind local compute---at an LLM inference cost under \$10/workload.\end{abstract}

\input{intro2}

\input{related}
\input{cufuse}
\input{eval}


\bibliographystyle{plainnat}
\bibliography{references.bib}

\appendix
\input{appendix/main}

\newpage
\input{checklist.tex}

\end{document}

%% file: header.tex
\newcommand{\cufuse}{\texttt{CUCo}}

\newcommand{\AnonymousCucoMirrorURL}{https://anonymous.4open.science/r/CUCo-06EC/}

\usepackage{xspace} \newcommand{\eg}{e.g.\xspace}
\usepackage{enumitem}
\usepackage{listings}
\usepackage{textcomp} 
\usepackage{pifont} 
\usepackage{algorithm}
\usepackage{algpseudocode}
\usepackage{capt-of}
\usepackage{afterpage}
\usepackage{hyphenat}

\newcommand{\cmark}{\ding{51}} 
\newcommand{\xmark}{\ding{55}} 

\definecolor{brewerpurple}{HTML}{AF4EA3}
\definecolor{brewerblue}{HTML}{377EB8}
\definecolor{NavyBlue}{HTML}{006EB8}
\definecolor{BrickRed}{HTML}{B6321C}
\definecolor{ForestGreen}{HTML}{009B55}
\lstdefinestyle{customc2}{
  emph={update\_metrics, prune\_completed\_jobs, exec\_jobs, accept, schedule, place, AcceptAll, Fifo, Consolidated, pop\_wait\_queue, update\_cluster},
  escapeinside={(*@}{@*)},
  emphstyle=\bfseries\color{NavyBlue},
  commentstyle=\color{ForestGreen}\itshape\ttfamily,
  morekeywords={yield},
  stringstyle=\color{red}\ttfamily,
  emph={[2]admission\_policy, scheduling\_policy, placement\_policy},emphstyle={[2]\bfseries\color{BrickRed}},
  language=Python,
  keywordstyle=\bfseries\color{green!40!black}
}

%% file: intro2.tex
\vspace{-8pt}
\section{Introduction}
\vspace{-8pt}

In distributed training and inference, communication sits squarely on the critical path alongside computation~\cite{megatron-lm, nanoflow}. To mitigate these communication bottlenecks, researchers have proposed techniques such as compression, overlap, and diverse forms of parallelism. However, recent systems like DeepEP~\cite{deepep2025} and Flux~\cite{chang2024flux} demonstrate that the latency of launching communication kernels has rapidly become a dominant overhead.

To overcome this limitation, GPU vendors have upended traditional host-driven communication—where the CPU launches all kernels—by enabling device-initiated execution. Interfaces such as NVSHMEM~\cite{nvshmem} and NCCL's device-side APIs~\cite{nccl_deviceapi} allow GPU threads to issue network requests directly from within a compute kernel, facilitating fine-grained interleaving at the tile or warp level. However, this architectural shift introduces a complex new design space. To effectively overlap computation and communication, developers must now make intricate decisions regarding the underlying communication backend, synchronization primitives, issuer granularity, and transfer placement. While expert-crafted libraries like DeepEP~\cite{deepep2025}, FLUX~\cite{chang2024flux}, and TokenWeave~\cite{gond2025tokenweave} manually navigate this space to achieve substantial speedups, their optimizations remain tightly coupled to specific workloads and network topologies.

\begin{figure*}[t]
\centering
\begin{minipage}[t]{0.48\textwidth}
    \centering
    \setlength{\abovecaptionskip}{2pt}
    \setlength{\belowcaptionskip}{-10pt}
    \includegraphics[width=\linewidth]{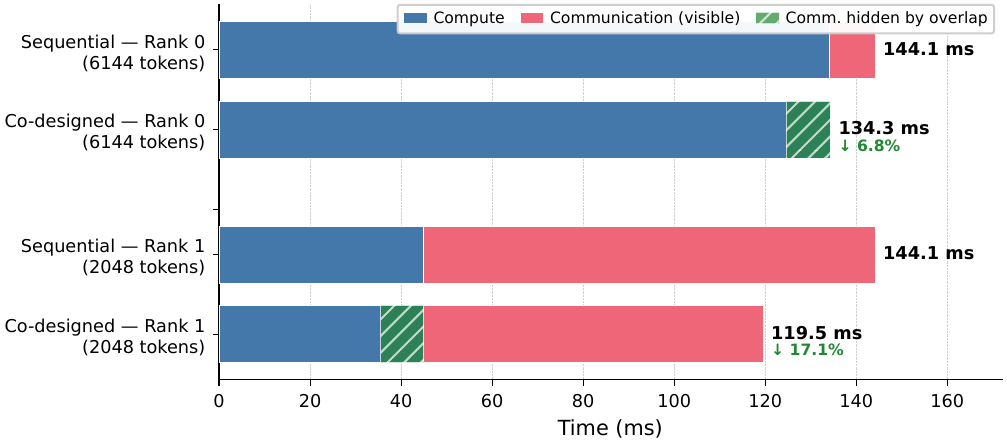}
    \caption{\small{\textbf{Why co-design matters.} Per-rank latency breakdown of a DeepSeek-V3 MoE layer (2 GPUs, inter-node).}}
    \label{fig:motivation}
\end{minipage}\hfill
\begin{minipage}[t]{0.48\textwidth}
    \centering
    \setlength{\abovecaptionskip}{2pt}
    \setlength{\belowcaptionskip}{-10pt}
    \includegraphics[width=\linewidth]{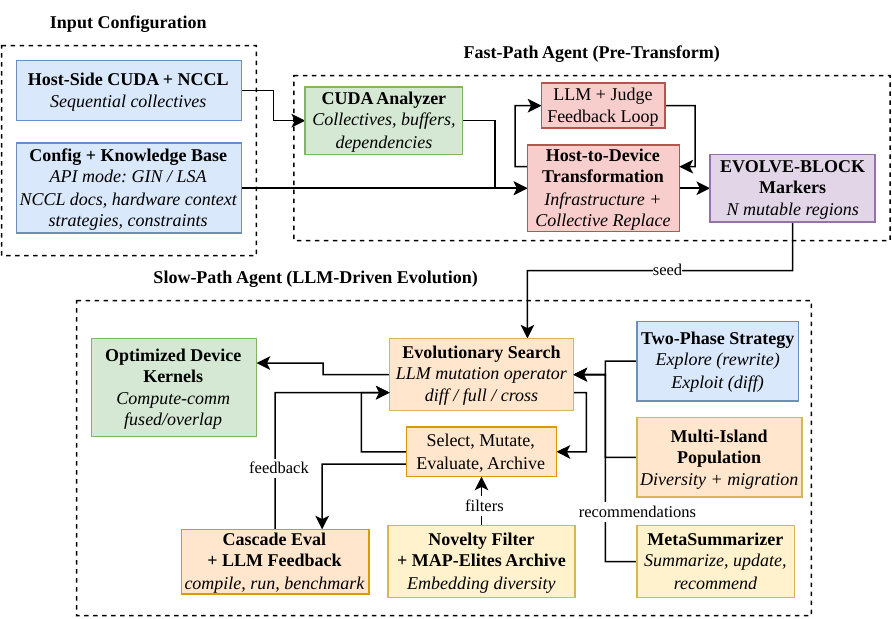}
    \caption{\small{Overall workflow of \cufuse{}.}}
    \label{fig:workflow}
\end{minipage}
\vspace{-17pt}
\end{figure*}


However, this design space is highly volatile. Evolving hardware generations, diverse interconnect topologies (such as NVLink, InfiniBand, RoCE, and PCIe), and emerging model architectures with novel communication patterns (\eg, Mixture-of-Experts all-to-all~\cite{deepep2025}) constantly shift the optimal design parameters. Consequently, maintaining efficiency across varying deployments requires frequent, manual kernel rewrites. Figure~\ref{fig:motivation} illustrates this concretely: during a DeepSeek-V3 MoE layer, communication consumes up to 69\% of the wall time on lightly-loaded ranks, yet a carefully co-designed strategy—overlapping dispatch with compute using variable-size transfers—can reduce end-to-end latency by 7--17\%. Unfortunately, modern ML compilers offer no relief. They assume communication occurs strictly outside the kernel boundary~\cite{triton, xla, zheng2025tilelink}; migrating it inside violates their foundational assumptions for operator scheduling and memory-access planning~\cite{10.1109/SC41406.2024.00094}. While bespoke, expert-written kernels achieve high performance by rigorously tuning for specific configurations, adapting them to new workloads demands navigating a massive design space of tiling, synchronization, and chunk sizing~\cite{chang2024flux, zheng2025tilelink}. This complexity restricts effective co-design to a small fraction of developers with deep systems expertise. Furthermore, these highly tuned stacks frequently rely on specific assumptions—such as the presence of RDMA or particular NVLink topologies—that are absent in many computing clusters, preventing practitioners from utilizing a single, drop-in library across heterogeneous environments~\cite{deepep2025, licker2025rdmap2p}.

However, new GPU generations and evolving interconnects like NVLink, Infiniband etc  change the choice of design parameters. New model structures introduce novel communication patterns (\eg Mixture-of-Experts all-to-all~\cite{deepep2025}), and deployment topologies vary across NVLink, InfiniBand, RoCE, and PCIe---each altering the co-design trade-offs and requiring kernel rewrites to maintain efficiency. Figure~\ref{fig:motivation} illustrates this concretely: on the DeepSeek-V3 MoE layer, communication consumes up to 69\% of wall time on the lightly-loaded rank, yet carefully co-designed strategies---overlapping dispatch with compute, using variable-size transfers---can reduce end-to-end latency by 7--17\%. Modern ML compilers offer no relief: they assume communication lives outside the kernel boundary~\cite{triton, xla, zheng2025tilelink}, and moving it inside violates the assumptions underlying operator scheduling and memory-access planning~\cite{10.1109/SC41406.2024.00094}. While bespoke, expert-written kernels are tuned for specific configurations---and that specificity is precisely what makes them fast---adapting them to new workloads and deployments requires balancing a large design space across tiling, synchronization, chunk sizing etc.~\cite{chang2024flux, zheng2025tilelink}, limiting co-design to developers with deep systems expertise. Moreover, existing stacks often assume NVLink, RDMA, GPU generations and specific topology configurations that are absent on many computing clusters, so practitioners cannot drop in the same library everywhere~\cite{deepep2025, licker2025rdmap2p}.

To address this bottleneck, we bring agentic AI to the co-design process. Given a host-driven baseline---where the CPU launches communication between compute kernels---our framework, \cufuse{}, automatically transforms it into a device-initiated kernel where GPU threads issue communication from within the compute code. The framework navigates the device-side design space (backend, placement, synchronization, granularity) to automatically discover high-performing configurations for each target topology.

\noindent\textbf{Target regime.} Co-design yields the largest gains when communication is on the critical path, when orchestration overhead is amplified across repeated exchanges (e.g., ring attention with many rounds) etc. Conversely, co-design is less likely to help when compute dominates wall time, when intra-node NVLink bandwidth makes communication negligible, or when an expert library already exposes the right fused schedule for the target workload and topology. \cufuse{} is designed for the former regime.

Naively prompting an LLM to produce co-designed kernels is inefficient, as it often generates candidates that fail to compile or deadlock at runtime. To make agentic co-design practical, the community needs a principled recipe. \cufuse{} provides one through three inter-twined components:
\begin{enumerate}[leftmargin=1.4em,itemsep=1pt,topsep=2pt]
  \item A \emph{structured design space} that constrains the agent's degrees of freedom to valid communication primitives, synchronization patterns, and tiling strategies---grounding reasoning in collective semantics rather than free-form code generation.
  \item A \emph{correctness-first fast path} that composes computation and communication into barrier-delimited phases, producing a reliable baseline before any optimization begins.
  \item A \emph{performance-driven slow path} that uses evolution-style search with empirical hardware feedback (latency etc.) to explore overlap strategies, chunk granularities, and stream topologies, converging on high-performance implementations tailored to each workload and hardware context.
\end{enumerate}

\vspace{-4pt}
A key insight underlying \cufuse{} is that expert communication libraries address a complementary layer of the problem. Libraries like DeepEP provide highly optimized communication primitives and even expose async APIs and multi-stream building blocks, yet the higher-level co-design decisions that determine end-to-end performance (overlap strategy, self/remote compute decomposition, stream topology, backend selection) are left to the user. \cufuse{} automates exactly this layer: for example, on intranode MoE dispatch, it discovers a two-stream overlap pipeline that hides dispatch behind local-chunk compute, a composition an expert could build manually on top of DeepEP's primitives.

Our contributions are as follows:
\vspace{-4pt}

\begin{enumerate}[leftmargin=1.4em,itemsep=1pt,topsep=2pt]
  \item \textbf{LLMs as bounded operators over domain-defined search spaces.} We formalize a structured configuration space for compute-communication co-design that constrains LLM agents to valid primitives, synchronization patterns, and tiling strategies. Existing expert systems~\cite{deepep2025, chang2024flux} can be expressed as points in this space. Without this structure, agents default to unconstrained code generation and repeatedly produce broken or suboptimal kernels.
  \item \textbf{Correctness-first decomposition for agentic optimization.} We introduce a fast-path/slow-path pipeline that separates correctness (constrained transformation to a verified seed) from performance (evolutionary search).
  \item \textbf{Portability through injected context, not retraining.} \cufuse{} adapts across intra-node (NVLink) and inter-node (InfiniBand) deployments by injecting hardware context into the agent prompt rather than baking topology knowledge into the system. This avoids the fine-tuning cost that limits prior agentic kernel generators~\cite{kevin} to fixed hardware targets.
\end{enumerate}

\cufuse{} is implemented in 18K lines of Python and evaluated on four multi-GPU workloads spanning Flash Attention, MoE dispatch/combine, KV-cache transfer, and GEMM+AllGather. Across these workloads, \cufuse{} achieves up to $1.57\times$ speedup over host-driven baselines. On a DeepSeek-V3 MoE layer, \cufuse{}'s discovered two-stream overlap strategy outperforms standard DeepEP usage by 12\%---not through faster primitives, but by hiding dispatch behind self-compute. All results are achieved at an LLM cost under \$10 per workload. To support the community, we release \cufuse{} as open source: \url{\AnonymousCucoMirrorURL}.

%% file: related.tex
\vspace{-15pt}
\section{Related Work}
\vspace{-3pt}
\begin{table}[t]
\centering
\scriptsize
\caption{Comparison of agentic CUDA kernel generation systems.}
\label{tab:agentic-cuda-comparison}
\vspace{-5pt}
\begin{tabular}{lcccc}
\toprule
\textbf{System} & \textbf{Multi-GPU} & \textbf{Comm Support} & \textbf{No Training} & \textbf{Device-Init} \\
\midrule
CUDAForge~\cite{cudaforge}      & \xmark & \xmark & \cmark & \xmark \\
STARK~\cite{dong2025starkstrategicteamagents}          & \xmark & \xmark & \cmark & \xmark \\
Kevin-32B~\cite{kevin}      & \xmark & \xmark & \xmark & \xmark \\
KernelFalcon~\cite{kernelfalcon}   & \xmark & \xmark & \cmark & \xmark \\
\textbf{\cufuse{}}         & \cmark & \cmark & \cmark & \cmark \\
\bottomrule
\end{tabular}
\vspace{-15pt}
\end{table}

\vspace{-8pt}
\subsection{Device-Initiated Communication}
\vspace{-8pt}
Traditional distributed GPU programs use \emph{host-driven} communication: the CPU launches collectives between compute kernels, limiting overlap to coarse-grained stream pipelining~\cite{megatron-lm, coconet}. Recent \emph{device-initiated} interfaces---NVSHMEM~\cite{nvshmem} and NCCL's device-side APIs~\cite{nccl_deviceapi}---allow GPU threads to issue communication from within a kernel. NCCL exposes two backends: GPU-Initiated Networking (GIN) for RDMA-style inter-node transfers and Load/Store Accessible (LSA) for direct peer-memory access over NVLink (Appendix~\ref{sec:nccl_demo}). This moves communication \emph{inside} the kernel boundary, enabling fine-grained interleaving at the tile or warp level.

\vspace{-3pt}
\subsection{Compute-Communication Co-design}
\vspace{-3pt}
Device-initiated APIs open a design space---backend selection, synchronization scope, issuer granularity, transfer placement---that determines how compute and communication overlap. Industrial stacks~\cite{deepep2025, chang2024flux, gond2025tokenweave, licker2025rdmap2p} navigate this space manually, achieving large speedups but requiring deep expertise and remaining tied to specific topologies. This is distinct from compute-only fusion (e.g., FlashAttention~\cite{dao2023flashattention2fasterattentionbetter}), which reduces memory traffic but keeps communication outside the kernel. \cufuse{} automates the device-initiated co-design process.

\vspace{-8pt}
\subsection{Agents for Kernel Generation}
\vspace{-3pt}
Recent agentic approaches~\cite{concur,cudaforge,evoengineer} combine an LLM with structured feedback to iteratively refine kernels, but fall short for distributed settings (Table~\ref{tab:agentic-cuda-comparison}): they lack collective primitives~\cite{cudaforge,evoengineer}, some require costly fine-tuning~\cite{kevin, ouyang2025kernelbenchllmswriteefficient} that is brittle to API changes, and they attempt direct code generation without decomposition~\cite{lange2025robustagenticcudakernel}---causing unrecoverable failures on complex compute-communication kernels where high-level coordination decisions (topology, primitive choice, synchronization granularity) must be resolved first.

\cufuse{} addresses these limitations by embedding multi-GPU semantics into a structured search space and employing a fast-path/slow-path decomposition.

%% file: cufuse.tex
\vspace{-10pt}
\section{CUCo}
\vspace{-10pt}
\noindent
{\bf Goals and Overview.}
\cufuse{} is an end-to-end agentic framework for generating fused compute-communication CUDA kernels (Figure~\ref{fig:workflow}). Why not simply prompt an LLM iteratively? Compute-communication co-design requires jointly resolving multiple interacting decisions---backend selection, synchronization strategy, overlap placement, issuer granularity---whose correctness constraints are subtle and topology-dependent. Without structured guidance, LLMs default to unconstrained code generation that either deadlocks at runtime or converges to local optima that leave performance on the table. Our ablations (Section~\ref{sec:ablation}) confirm this: naive iterative prompting yields only marginal improvements on the host baselines, while skipping a correctness-first stage wastes the search budget on broken candidates. These observations motivate the three components below:

(i)~\emph{Design space specification}---a structured configuration space $\mathcal{C}$ so agents reason over device-initiated communication and overlap using minimal user-specified detail rather than unconstrained code generation;
(ii)~\emph{Fast-path agent}---correctness-first transformation from host-driven baselines to verified device-initiated kernels across backends and schedules;
(iii)~\emph{Slow-path agent}---performance-driven search over overlap strategies, pipelining, and placement within $\mathcal{C}$ to discover high-performance kernels.

\vspace{-10pt}
\subsection{Design Space Specification}
\vspace{-8pt}
\label{subsec:design_space}

Transitioning communication into the kernel shifts the paradigm from lockstep host-driven collectives to thread-driven device-initiated transfers. This introduces a complex configuration problem across multiple interacting dimensions. Without explicit guidance, an LLM agent makes these decisions implicitly from training priors, often producing suboptimal or incorrect kernels (our ablation in Section~\ref{sec:ablation} confirms this).

To prevent this, \cufuse{} formalizes an \emph{optimization configuration space}, defined by eight discrete dimensions: $\mathcal{C} = \mathcal{B} \times \mathcal{M} \times \mathcal{P} \times \mathcal{S} \times \mathcal{I} \times \mathcal{G} \times \mathcal{O} \times \mathcal{K}$, representing backend ($\mathcal{B}$), completion mechanism ($\mathcal{M}$), placement ($\mathcal{P}$), synchronization scope ($\mathcal{S}$), issuer granularity ($\mathcal{I}$), chunk size ($\mathcal{G}$), memory ordering ($\mathcal{O}$), and context multiplicity ($\mathcal{K}$). These dimensions capture the dominant degrees of freedom for compute-communication co-design, as evidenced by their ability to express existing expert-crafted systems (Table~\ref{tab:design_space_coverage}). While some parameters dictate raw transport ($\mathcal{B}$), others expose subtle architectural trade-offs. For instance, the synchronization mechanism ($\mathcal{M}$) dictates how completion is detected: barriers impose global rendezvous, whereas signals enable point-to-point polling for tighter overlap but introduce race-condition risks. Similarly, communication placement ($\mathcal{P}$) determines whether transfers are deferred, tile-fused within a persistent kernel, or pipelined across streams. The best choice is highly topology-dependent: a synchronization scope ($\mathcal{S}$) of \texttt{World} may be required across slower inter-node links, whereas \texttt{Local} or \texttt{Rail} scopes dominate on dense fabrics (GIN issues RDMA puts over the network; LSA performs direct load/store to peer memory over NVLink). Table~\ref{tab:fusion_space} and Table~\ref{tab:design_space_coverage} detail this space and demonstrate how expert-crafted systems map onto these axes (see Appendix~\ref{sec:design_space_details} for a full semantic breakdown).

\begin{table}[t]
\centering
\scriptsize
\caption{Optimization directive dimensions. Concrete dimensions correspond to fixed NCCL API identifiers. Intent-based dimensions express high-level objectives that the agent interprets during code generation.}
\begin{tabular}{p{0.3\columnwidth} p{0.65\columnwidth}}
\toprule
\textbf{Concrete API Dimensions} & \textbf{Allowed Values} \\
\midrule
$\mathcal{B}$: Backend     & \texttt{GIN} \;|\; \texttt{LSA} \;|\; \texttt{Hybrid} \\
$\mathcal{M}$: Sync Mechanism & \texttt{Barrier} \;|\; \texttt{Signal} \;|\; \texttt{SignalShadow} \;|\; \texttt{Counter} \\
$\mathcal{I}$: Issuer      & \texttt{Thread} \;|\; \texttt{Warp} \;|\; \texttt{WarpSpan} \;|\; \texttt{Tile\textlangle N\textrangle} \;|\; \texttt{CTA} \\
$\mathcal{O}$: Ordering    & \texttt{Relaxed} \;|\; \texttt{Acquire} \;|\; \texttt{Release} \;|\; \texttt{AcqRel} \\
$\mathcal{K}$: Contexts    & $1$ \;|\; $2$ \;|\; $N$ \\
\midrule
\textbf{Intent-Based Dimensions} & \textbf{Allowed Values} \\
\midrule
$\mathcal{P}$: Placement     & Deferred \;|\; Tile-fused \;|\; Tile-pipelined \;|\; Stream-split \\
$\mathcal{S}$: Sync Scope    & Local \;|\; World \;|\; Rail \;|\; Hierarchical \\
$\mathcal{G}$: Granularity   & Per-peer \;|\; Per-tile \;|\; Per-chunk \\
\bottomrule
\end{tabular}
\label{tab:fusion_space}
\vspace{-10pt}
\end{table}

\begin{table}[t]
\centering
\scriptsize
\caption{Expert-crafted systems mapped onto the design space $\mathcal{C}$ by semantic equivalence. These systems use custom transport implementations, not NCCL's device API. DeepEP's IB path uses custom RDMA one-sided puts (GIN semantics); its NVLink path uses direct peer memory stores (LSA semantics).}
\label{tab:design_space_coverage}
\begin{tabular}{lcccccccc}
\toprule
\textbf{System} & $\mathcal{B}$ & $\mathcal{M}$ & $\mathcal{P}$ & $\mathcal{S}$ & $\mathcal{I}$ & $\mathcal{G}$ & $\mathcal{O}$ & $\mathcal{K}$ \\
\midrule
DeepEP (NVL)  & LSA    & Barrier      & Deferred       & Local & CTA  & Per-peer & Release & -- \\
DeepEP (IB)   & GIN    & Signal       & Deferred       & World & CTA  & Per-peer & Acquire & 1 \\
FLUX          & LSA    & Barrier      & Tile-fused     & Local & Warp & Per-tile & AcqRel  & -- \\
\bottomrule
\end{tabular}
\vspace{-25pt}
\end{table}

\noindent\textbf{Optimization Directive.}
To use $\mathcal{C}$ as an agent reasoning interface, \cufuse{} requires each agent to emit an explicit \emph{optimization directive} before generating any kernel code. The directive selects one value per dimension of $\mathcal{C}$ (Appendix~\ref{sec:directive_listing}). The fast-path agent always emits a fixed conservative directive (CTA issuance, barrier completion, world scope, release ordering, single context). The slow-path agent treats the directive as the search variable, reasoning over $\mathcal{C}$ to emit progressively more aggressive configurations for a given workload and hardware target. By requiring an explicit directive before any code is written, \cufuse{} makes each agent's design decisions inspectable and directly traceable to the kernels they produce.
\noindent\textbf{Agent Context.}
General-purpose LLMs have almost no exposure to NCCL's device-initiated API: it is recent, sparsely documented, and heavily conflated with host-driven semantics in the LLM's pretraining data. To prevent hallucinated API calls, \cufuse{} dynamically injects a structured context into the prompt, \emph{strictly conditioned on the selected backend} $\mathcal{B}$ (e.g., if GIN is selected, only GIN documentation is injected; LSA material is omitted entirely). The context comprises three components:
(1)~\emph{API Documentation}: Schema-driven interface specs, semantic preconditions, and a compilable reference kernel.
(2)~\emph{Strategy Knowledge}: Best-practice overlap strategies (kernel-level fusion vs.\ multi-stream pipelining) alongside backend-specific correctness rules to constrain the search space of plausible-but-incorrect programs.
(3)~\emph{Hardware Context}: Dynamically extracted GPU properties, network topologies, and SM resource limits, ensuring that agent decisions reflect the target hardware rather than training priors.
Full details of the agent context, including an LSA interface sample, are provided in Appendix~\ref{sec:agent_context}.

\vspace{-10pt}
\subsection{Fast-Path Agent}
\vspace{-8pt}
\label{fast_path}
The fast‑path agent is dedicated entirely to producing a correct device-initiated baseline, prioritizing functional verification over performance (Figure~\ref{fig:workflow}). Attempting to generate a fully optimized kernel from scratch leads to high failure rates (our ablation shows $>$25\% of the evolution budget wasted on broken candidates that fail compilation or deadlock). Instead, \cufuse{} utilizes a three-step correctness-first pipeline: (1)~\emph{CUDA Code Analysis}: A static analyzer extracts the communication dependency graph from the host-driven baseline, recovering the precise compute--communication boundary (Appendix~\ref{app:analysis_example}). (2)~\emph{Host-to-Device Transformation}: The agent uses a two-stage LLM--judge feedback loop to rewrite host collectives into device-initiated equivalents. \emph{Stage~A} sets up the device-communication infrastructure (symmetric memory allocation, window registration, device communicator instantiation). \emph{Stage~B} replaces host collectives with device primitives (\texttt{put}/\texttt{waitSignal} for GIN; peer stores with barrier synchronization for LSA) using the fixed conservative directive described above to ensure deterministic execution. (3)~\emph{Evolve-Block Annotation}: The verified program is annotated with paired markers identifying mutable regions. Frozen regions (initialization, verification) are excluded to prevent downstream mutations from breaking the evaluation harness.

Separating infrastructure setup from semantic replacement isolates LLM generation tasks, reducing the failure space and allowing the LLM judge to provide targeted feedback on compilation and logic errors. The resulting verified program becomes the \emph{seed} for the slow-path agent. Full details of the fast-path transformation are provided in Appendix~\ref{sec:fast_path_details}.

\vspace{-10pt}
\subsection{Slow-Path Agent}
\vspace{-10pt}
\label{subsec:slow_path}

The slow‑path agent navigates the search space $\mathcal{C}$ using an island‑based evolutionary optimizer augmented with an LLM mutation operator (Appendix~\ref{sec:slowpath_algo}). Unlike standard genetic algorithms where random perturbation ruins code validity, \cufuse{} uses an LLM to propose semantically informed edits bounded by the directive space $\mathcal{C}$ and spatial \texttt{evolve-block} markers. The framework explicitly schedules exploration vs.\ exploitation through a phase-dependent mutation policy (Appendix~\ref{sec:llmmutate_details}) and prevents premature convergence by evolving multiple independent islands~\cite{lange2025shinkaevolveopenendedsampleefficientprogram} that periodically exchange high-scoring candidates (Appendix~\ref{sec:multi-island}).

\noindent\textbf{Explore-Exploit Mutation Policy.} The search relies on three LLM mutation modes: \emph{full rewrites} (large-step architectural changes), \emph{diff patches} (fine-grained parameter tuning), and \emph{cross-pollination} (crossover synthesis from a MAP-Elites~\cite{mapelites} archive). Because exhaustive exploration is wasteful, \cufuse{} explicitly schedules these three mutation modes. The \emph{Explore phase} favors high-temperature full rewrites to discover diverse baseline architectures (e.g., fused kernels vs.\ multi-stream pipelining) before any single design dominates. The \emph{Exploit phase} transitions to low-temperature diff patches, greedily refining synchronization scopes, launch bounds, and memory orderings of the most promising candidates.

\noindent\textbf{Cascade Evaluation \& Feedback Routing.}
Every generated offspring undergoes a fast-fail three-level evaluation cascade: compile ($\ell_1$), numerically verify via \texttt{mpirun} ($\ell_2$), and benchmark latency ($\ell_3$). Results are scored and persisted in an embedding-guided Candidate Database. A specialized LLM feedback agent analyzes each termination trace (compiler errors, mismatches, or performance metrics) and generates a diagnostic. When selecting the next parent for mutation, the prompt is populated with: (1)~\emph{Parent history} (source code, metrics, and targeted feedback), (2)~\emph{Structurally similar peers} (retrieved via code embeddings as negative examples to prune failed paths), and (3)~\emph{Meta-recommendations} (cross-generation strategic trends from a meta-summarizer, Appendix~\ref{sec:meta}). This closed-loop feedback drives continuous improvement across generations.

\par

%% file: eval.tex
\vspace{-10pt}
\section{Evaluation}
\vspace{-10pt}
Our experiments address three questions: (i)~Does \cufuse{} improve latency over host-orchestrated NCCL on diverse compute--communication kernels? (ii)~How does automated co-design compare to an expert-tuned stack (DeepEP, Section~\ref{sec:deepep-comparison})? (iii)~Which pipeline components matter (Section~\ref{sec:ablation})?

The primary quantitative comparisons are \cufuse{} vs.\ host-driven baselines that mirror realistic user code: CPU-launched NCCL collectives and streams as typically written before device-initiated redesign. We implement these host paths with strong conventional overlap---multi-stream scheduling, established collective patterns, and padding only where required---reflecting the most efficient NCCL-centric usage for each workload.


\noindent\textbf{Benchmarks and workload scope.}
There is, to our knowledge, no public benchmark or dataset of \emph{distributed} multi-GPU programs that jointly exercises computation and communication co-design across heterogeneous topologies. Kernel-generation benchmarks such as KernelBench~\cite{ouyang2025kernelbenchllmswriteefficient} focus on single-GPU tasks rather than multi-GPU programs under varied interconnects. We therefore evaluate on four workloads we construct to cover dominant LLM-side patterns: ring-style attention with repeated peer exchanges (long-context parallelism); MoE dispatch/combine with All-to-All traffic and skewed expert routing; KV handoff for disaggregated prefill--decode serving; and GEMM~+~AllGather as a minimal post-compute collective. Together they span ring, All-to-All, point-to-point handoff, and AllGather semantics.

\noindent\textbf{Environment Setup.} Experiments were conducted across two servers, each featuring dual 16-core Intel Xeons, 256 GiB RAM, and 4x 80 GiB NVIDIA A100 GPUs. Intra-node communication relies on NVLink, and inter-node communication uses RoCE. The system runs Ubuntu 22.04.4 (Linux 5.15), utilizing CUDA 13.1 and NCCL 2.28.9. All LLM agents use Anthropic's Claude Sonnet 4.5.

\noindent\textbf{Search Cost.} For each workload, the slow-path agent evaluates candidates through the three-level cascade over a fixed budget. Across all workloads, the fast-path required 2--3 iterations and the slow-path 10--20 generations, taking 0.8--1.8 hours of wall time and costing \$4.98--\$10.96 in LLM API fees---far less than the hardware cost of the workloads being optimized. A per-workload breakdown is provided in Table~\ref{tab:search-cost} (Appendix~\ref{sec:search-cost-details}).

\begin{figure*}[t]
\centering
\setlength{\abovecaptionskip}{-2pt}
\setlength{\belowcaptionskip}{-8pt}
\begin{minipage}[t]{0.24\textwidth}
    \centering
    \includegraphics[width=\linewidth]{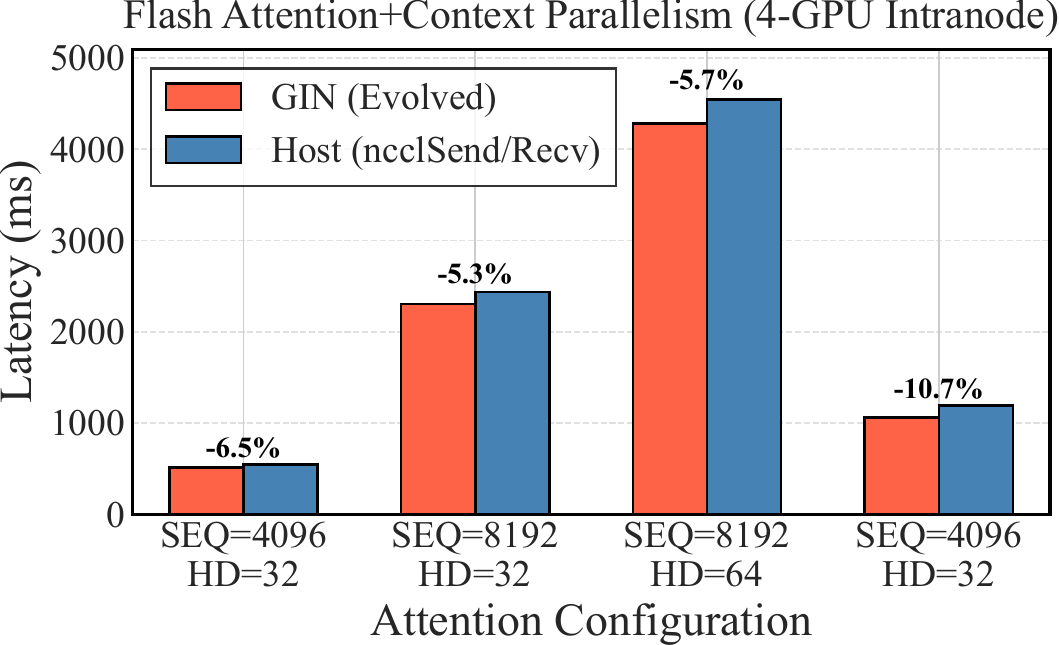}
    \captionof{figure}{\small{Flash Attention with Context Parallelism with varying sequence length and attention head dimension.}}
    \label{fig:flash-attn}
\end{minipage}
\hfill
\begin{minipage}[t]{0.24\textwidth}
    \centering
    \includegraphics[width=\linewidth]{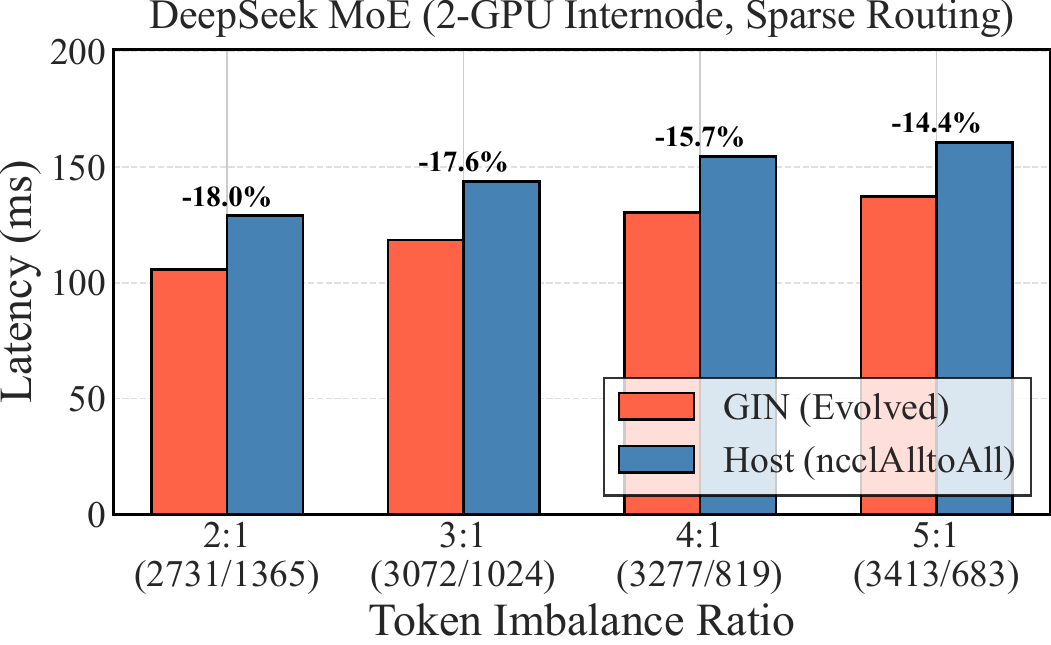}
    \captionof{figure}{\small{DeepSeek-V3 MoE layer across inter-node RoCE links with expert skewness.}}
    \label{fig:moe}
\end{minipage}
\hfill
\begin{minipage}[t]{0.24\textwidth}
    \centering
    \includegraphics[width=\linewidth]{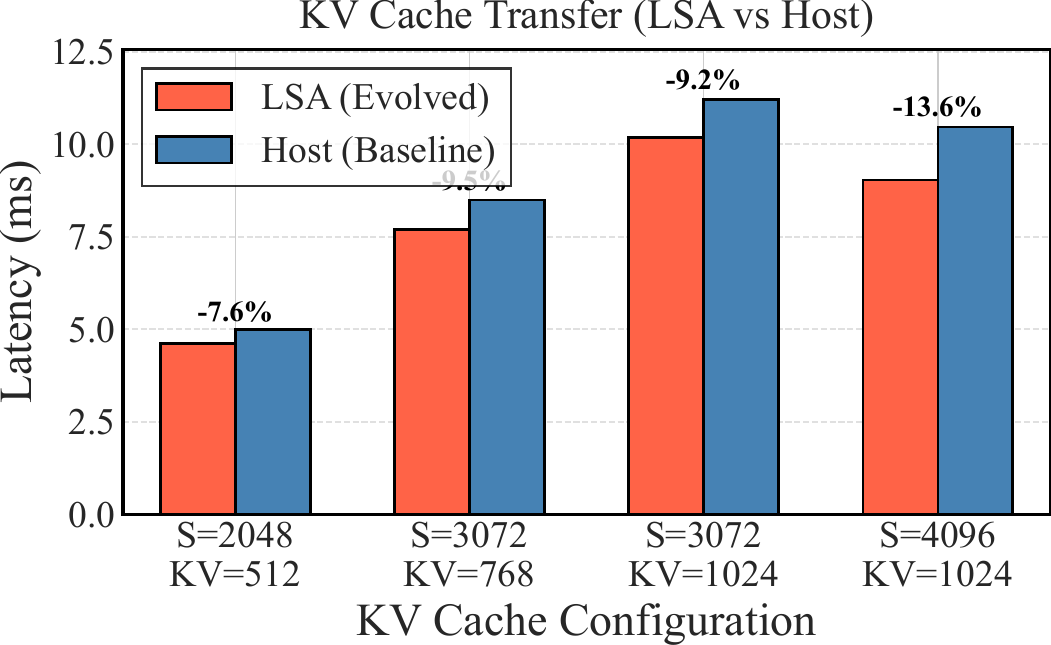}
    \captionof{figure}{\small{Intra-node KV cache transfer latency across varying sequence lengths and KV dimensions.}}
    \label{fig:kv-cache}
\end{minipage}
\hfill
\begin{minipage}[t]{0.24\textwidth}
    \centering
    \includegraphics[width=\linewidth]{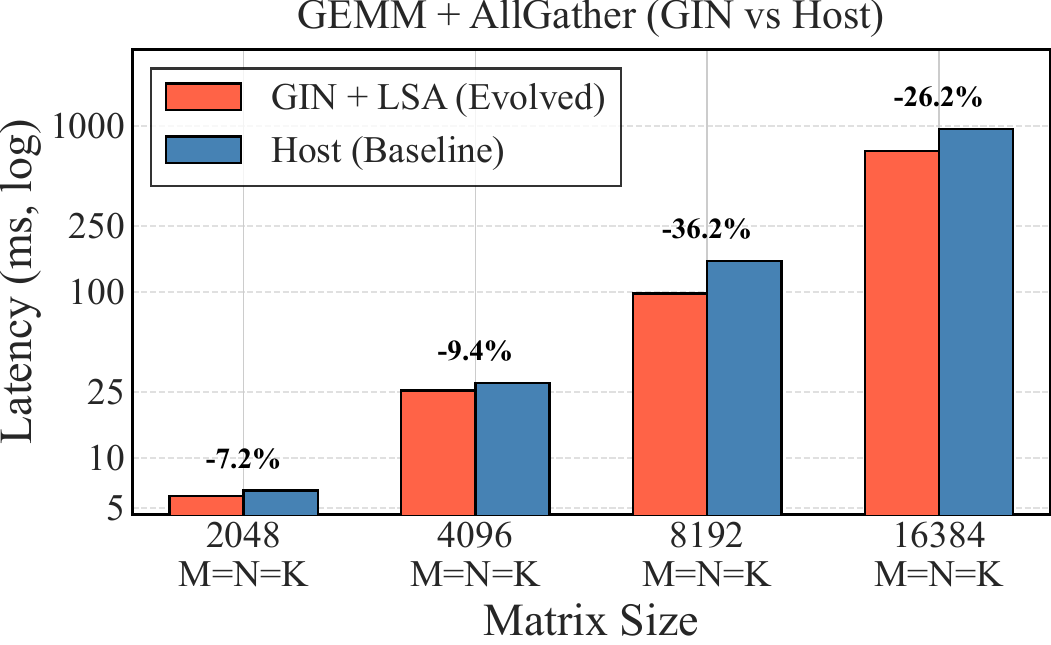}
    \captionof{figure}{\small{Intra- and inter-node GEMM + AllGather latency across square matrix sizes.}}
    \label{fig:gemm-allgather}
\end{minipage}
\end{figure*}

\noindent\textbf{Workloads.} We evaluate on four workloads that cover dominant LLM communication patterns; Figures~\ref{fig:flash-attn}--\ref{fig:gemm-allgather} summarize end-to-end results. Table~\ref{tab:strategy} previews the co-design strategy \cufuse{} discovered for each workload and the design-space choices that realize it.

\noindent\underline{\emph{Flash Attention with Context Parallelism.}}
This workload implements Ring Attention~\cite{liu2024ringattention}, where each GPU owns one Q shard while KV shards rotate. The bottleneck is inter-round pipeline bubbles. We evaluate on 4 intra-node GPUs (NVLink) with $\text{SEQ} \in \{4096, 8192\}$ and $\text{HD} \in \{32, 64\}$.

\noindent\underline{\emph{DeepSeek-V3 MoE Dispatch and Combine.}}
Each GPU routes tokens to remote experts via AlltoAll dispatch, runs expert GEMMs, and gathers results via a reverse AlltoAll combine. The bottleneck is sequential comm-compute phases and padding overhead under imbalanced routing. We evaluate on 2~GPUs across inter-node RoCE links with 4096~tokens per rank under four skew ratios (2:1 to 5:1).

\noindent\underline{\emph{KV-Cache Transfer.}}
In disaggregated prefill-decode serving, the prefill GPU must compute and send K and V projections before decode can begin. The bottleneck is the compute-to-send idle gap from CPU-launched transfers. We evaluate on two NVLink-connected GPUs with hidden dimension 4096.

\noindent\underline{\emph{GEMM + AllGather.}}
Each rank performs a local GEMM and then participates in an AllGather to distribute the full output. This isolates the simplest post-compute collective with no cross-rank dependencies. We evaluate on 4x GPUs spanning intra-node (NVLink) and inter-node (RoCE) links.

\noindent Full workload details, host baselines and CUCo-discovered pipelines, are provided in Appendix~\ref{sec:workload_details}.

\begin{table*}[t]
\centering
\scriptsize
\setlength{\tabcolsep}{4pt}
\caption{What \cufuse{}'s agent discovered: per-workload dominant bottleneck, the co-design strategy the evolutionary search converged on, and the key design-space choices ($\mathcal{C}$) that realize it.}
\label{tab:strategy}
\begin{tabular}{@{}p{0.10\textwidth}p{0.24\textwidth}p{0.37\textwidth}p{0.20\textwidth}@{}}
\toprule
\textbf{Workload} & \textbf{Dominant bottleneck} & \textbf{Discovered strategy} & \textbf{Key design choices} \\
\midrule
Flash Attention & Pipeline bubbles between ring exchange rounds & Persistent cooperative kernel with per-tile pipelining; comm block polls tile readiness via atomics & $\mathcal{P}$: tile-fused; $\mathcal{M}$: counter; $\mathcal{I}$: CTA (1 comm + 80 compute) \\
MoE Dispatch/ Combine & Sequential comm--compute phases; padding overhead on imbalanced ranks & Self/remote compute split with two-stream dispatch overlap; variable-size per-peer transfers & $\mathcal{B}$: GIN (inter-node); $\mathcal{P}$: stream-split; $\mathcal{G}$: per-peer \\
KV-Cache Transfer & Compute-to-send idle gap from host-launched sends & GPU-triggered chained send: K GEMM $\to$ send K $\to$ V GEMM $\to$ send V with signal & $\mathcal{I}$: device; $\mathcal{O}$: release; $\mathcal{M}$: signal \\
GEMM + AllGather & Post-compute idle before collective & Immediate device-initiated broadcast fused after GEMM completion & $\mathcal{P}$: deferred; $\mathcal{B}$: LSA (intra) / GIN (inter) \\
\bottomrule
\end{tabular}
\vspace{-13pt}
\end{table*}

\begin{figure*}[t]
\centering
\setlength{\abovecaptionskip}{0pt}
\setlength{\belowcaptionskip}{-10pt}
\begin{minipage}[t]{0.44\textwidth}
    \centering
    \includegraphics[width=\linewidth]{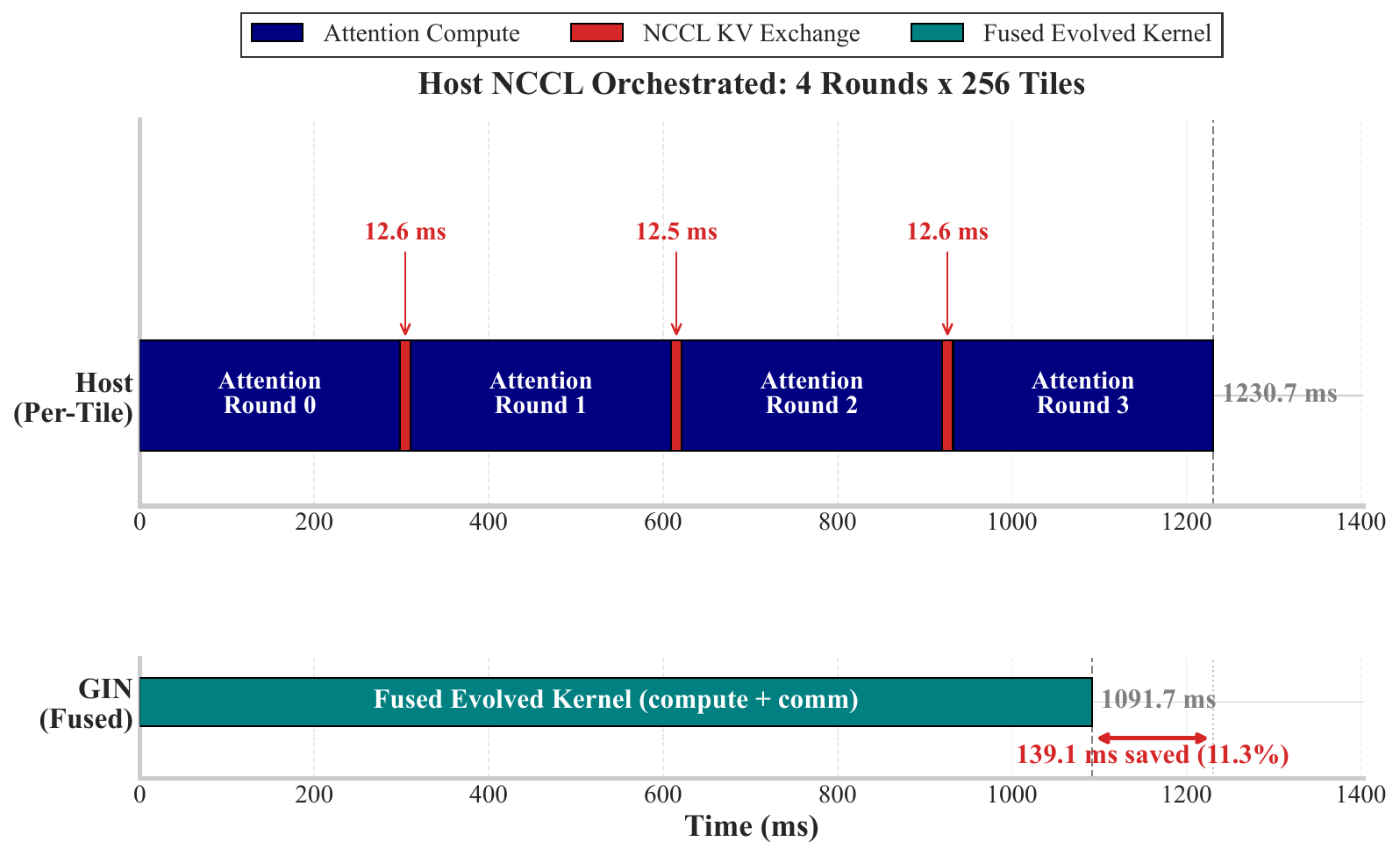}
    \captionof{figure}{\small{Flash Attention timeline: host NCCL baseline vs.\ \cufuse{} fused kernel ($\text{SEQ}=4096$, $\text{HD}=32$, 4~GPUs, NVLink). }}
    \label{fig:flash-attn-timeline}
\end{minipage}\hfill
\begin{minipage}[t]{0.55\textwidth}
    \centering
    \includegraphics[width=\linewidth]{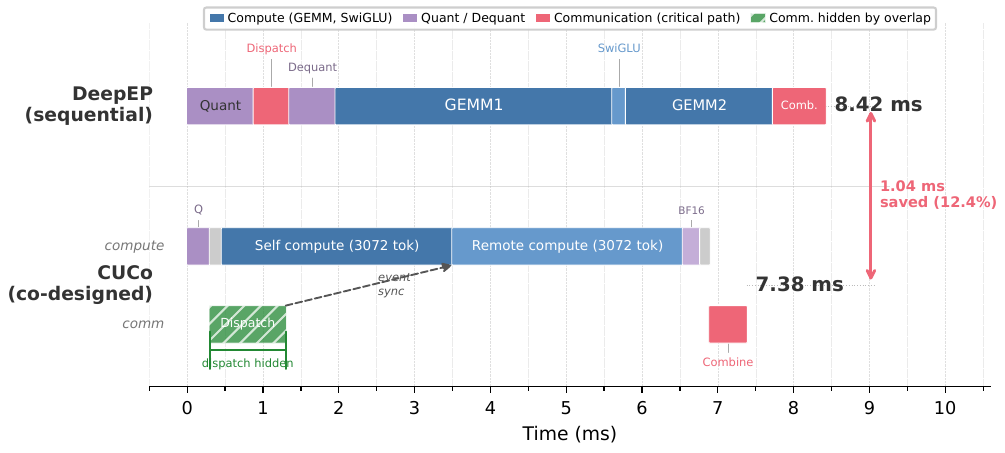}
\captionof{figure}{\small{\textbf{DeepEP vs.\ \cufuse{}} on the MoE layer (Rank~0, 6144 tokens). DeepEP runs phases sequentially. \cufuse{} hides dispatch behind self-compute via a two-stream split.}}
    \label{fig:deepep-timeline}
\end{minipage}
\vspace{-6pt}
\end{figure*}

\vspace{-3pt}
\subsection{\cufuse{}'s End-to-End Evaluation}
\vspace{-3pt}
\label{sec:e2e}

Figures~\ref{fig:flash-attn}--\ref{fig:gemm-allgather} report end-to-end latency for all four workloads. We compare \cufuse{}-evolved device-initiated CUDA against the host-driven NCCL baseline described above; latency is the median over five stable runs measured with CUDA events. Across the reported configurations, \cufuse{} lowers latency by 5.3\%--36.2\% relative to that baseline; our best reported configuration reaches up to $1.57\times$ speedup over host-driven NCCL, consistent with Section~1 and the abstract. The gains stem from fine-grained overlap (e.g., per-tile pipelining, dispatch hiding) that masks network latency behind computation, achieving tight interleavings impossible under host-level orchestration.

\vspace{-3pt}
\subsection{Case Study: Flash Attention with Context Parallelism}
\vspace{-3pt}
\label{sec:flash-attn-deep-dive}

The Flash Attention workload provides a clear demonstration of \cufuse{}'s architectural advantage because the ring structure amplifies per-round overheads across multiple exchange steps. We focus on the configuration ($\text{SEQ}=4096$, $\text{HD}=32$), which exhibits the largest speedup (\textbf{11.3\%}), and use Nsight Systems profiling to dissect where the gains originate. Figure~\ref{fig:flash-attn-timeline} shows the execution timelines.

The \cufuse{} evolved kernel launches a single cooperative grid with a dedicated \emph{communication block} that runs the entire ring protocol (GIN put/wait/flush across all exchange rounds) and separate \emph{compute blocks} that process attention tiles. The critical capability is \emph{per-tile pipelining}: as the comm block completes the transfer of tile~$j$, compute blocks immediately begin processing it---even while later tiles are still in flight. Host orchestration cannot achieve this: because the attention kernel saturates all SMs, a signaling kernel on the communication stream can never be scheduled, creating a priority inversion that forces purely sequential execution between exchange rounds.

As shown in Figure~\ref{fig:flash-attn-timeline}, per-tile pipelining contributes the largest savings (63.7\,ms), followed by eliminating compute-stream idle time during exchange rounds (37.7\,ms) and removing host proxy overhead (37.6\,ms). Together, these reduce end-to-end latency from \textbf{1230.7\,ms} to \textbf{1091.7\,ms}. Full architectural details (block counts, double-buffering, synchronization primitives) are in Appendix~\ref{sec:flash_attn_details}.

\vspace{-3pt}
\subsection{Comparison with Expert-Crafted Libraries}
\vspace{-3pt}
\label{sec:deepep-comparison}

\begin{figure*}[t]
\begin{minipage}[c]{0.58\textwidth}
\centering
\scriptsize
\setlength{\tabcolsep}{4pt}
\captionof{table}{Per-phase latency: DeepEP vs.\ \cufuse{} on the DeepSeek-V3 MoE layer (2~A100s, NVLink, Rank~0, 6144 tokens).}
\vspace{4pt}
\label{tab:deepep-comparison}
\begin{tabular}{lrrr}
\toprule
\textbf{Phase} & \textbf{DeepEP (ms)} & \textbf{\cufuse{} (ms)} & $\boldsymbol{\Delta}$ \textbf{(ms)} \\
\midrule
FP8 quantize              & 0.88 & 0.30 & $-$0.58 \\
Dispatch (comm)            & 0.47 & ${\sim}$1.0\textsuperscript{$\dagger$} & \emph{hidden} \\
Compute\textsuperscript{$\ddagger$} & 6.38 & 6.08 & $-$0.30 \\
BF16 cast + overhead       & ---  & 0.30 & +0.30 \\
Combine (comm)             & 0.68 & ${\sim}$0.70 & +0.02 \\
\midrule
\textbf{Total (wall clock)} & \textbf{8.42} & \textbf{7.38} & \textbf{$-$1.04} \\
\textbf{Dispatch on crit.\ path} & \textbf{yes} & \textbf{no} & \textbf{$-$12.4\%} \\
\bottomrule
\end{tabular}
\\[2pt]
{\scriptsize $\dagger$~Concurrent with self-compute; hidden behind 3.04\,ms of local-chunk work.}\\
{\scriptsize $\ddagger$~DeepEP: one GEMM (6144 tok). \cufuse{}: two GEMMs (3072+3072), same cuBLAS TF32.}
\end{minipage}\hfill
\begin{minipage}[c]{0.38\textwidth}
\centering
\includegraphics[width=0.92\linewidth]{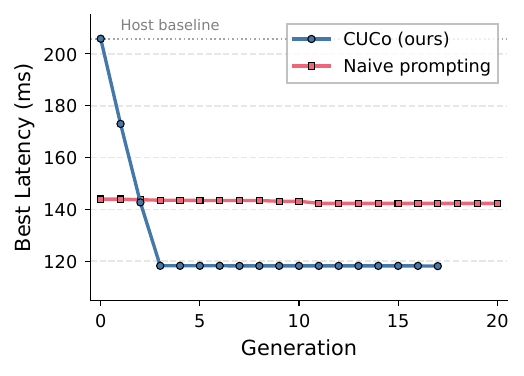}
\vspace{-4pt}
\captionof{figure}{\small{CUCo vs.\ naive iterative prompting on the GIN MoE workload. CUCo reaches 118\,ms by gen.~3; naive prompting plateaus at 142\,ms.}}
\label{fig:ablation-naive}
\end{minipage}
\vspace{-10pt}
\end{figure*}

\noindent\textbf{Why DeepEP (and not every expert stack).}
Among recent industrial stacks~\cite{deepep2025, chang2024flux, licker2025rdmap2p}, DeepEP was the only library we could install, run, and profile end-to-end on our hardware (A100, NVLink intra-node, RoCE inter-node). It ships optimized transports and exposes async APIs (\texttt{async\_finish}, \texttt{previous\_event}, dedicated comm stream) that could enable overlap---but as a \emph{communication primitive library}, it leaves higher-level co-design decisions to the caller, precisely the layer \cufuse{} automates. We evaluate on the DeepSeek-V3 MoE layer (2 A100 GPUs, intranode NVLink, 4096 tokens, TF32 compute) using DeepEP's standard sequential single-stream flow. Table~\ref{tab:deepep-comparison} shows the per-phase breakdown.

\noindent\textbf{What \cufuse{} discovered.}
Figure~\ref{fig:deepep-timeline} visualizes the two execution models. The agent's evolutionary search converged on a two-stream overlap pipeline with three co-design decisions: (1)~\emph{self/remote compute split}---recognizing that tokens routed to the local expert can be processed immediately without waiting for the network, the agent splits the single GEMM into two smaller GEMMs (3072 tokens each); (2)~\emph{dispatch hiding}---running the dispatch kernel on a high-priority \texttt{comm\_stream} concurrently with self-chunk compute on a normal-priority \texttt{compute\_stream}, gating remote-chunk compute on dispatch completion via a CUDA event; and (3)~\emph{backend selection}---choosing LSA (direct peer memory stores) over GIN (RDMA-style put/signal) for the NVLink topology. Each decision is a point in the design space $\mathcal{C}$ (Table~\ref{tab:fusion_space}).

\noindent\textbf{Why \cufuse{} wins despite slower raw communication.}
DeepEP's primitives are faster in isolation (0.47\,ms vs.\ ${\sim}$1.0\,ms), but this is irrelevant when dispatch is hidden behind self-compute (3.04\,ms $>$ 1.0\,ms). The 1.04\,ms end-to-end saving corresponds to the dispatch plus quantization overhead that DeepEP exposes sequentially.

\noindent\textbf{Multi-layer scaling.}
Over $L{=}1$ to $16$ consecutive MoE layers, \cufuse{} maintains a stable 12--14\% per-layer advantage (7.29--7.38\,ms vs.\ 8.45--8.50\,ms), confirming that the overlap benefit is structural and per-layer, not a warmup artifact.


\vspace{-10pt}
\subsection{Ablation Studies}
\vspace{-10pt}
\label{sec:ablation}
\noindent\textbf{Fast-Path Agent.}
We evaluate whether the fast-path agent is necessary by comparing the full \cufuse{} pipeline (fast-path + slow-path) against a \emph{slow-path--only} ablation that runs the evolutionary agent directly on the original host NCCL code, bypassing the fast-path transformation entirely. Both configurations use the DeepSeek-V3 MoE workload (2-GPU inter-node, RoCE) with identical LLM models, evaluation harness, and evolutionary budgets.

\begin{figure*}[t]
\centering
\setlength{\abovecaptionskip}{2pt}
\setlength{\belowcaptionskip}{-8pt}
\begin{minipage}[t]{0.24\textwidth}
    \centering
    \includegraphics[width=\linewidth]{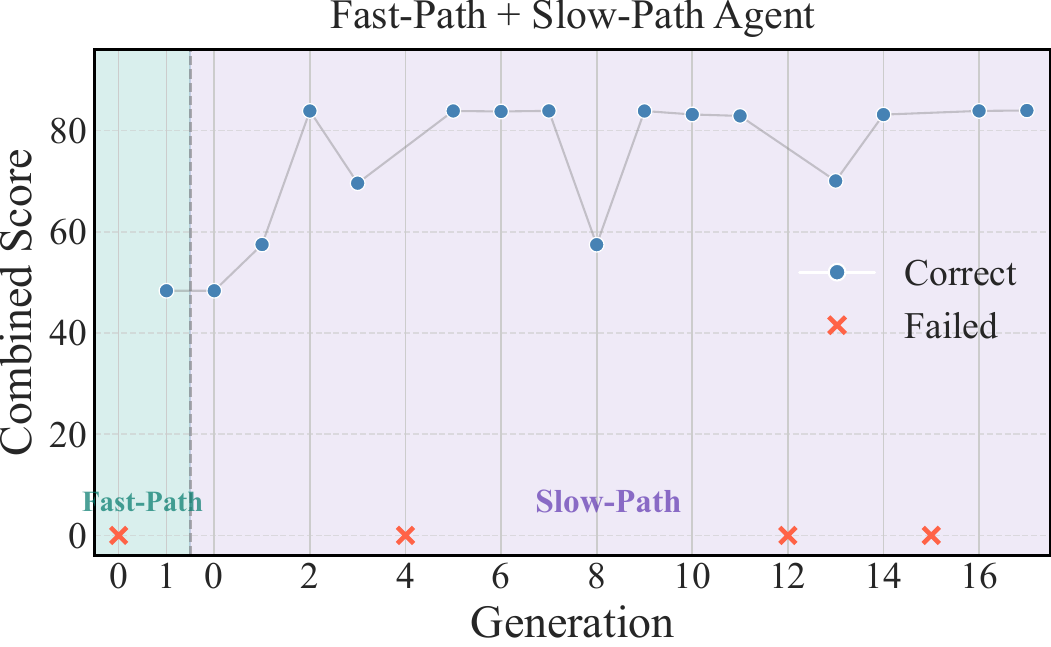}
    \captionof{figure}{\small{Fast-path + slow-path agent on the GIN MoE workload.}}
    \label{fig:ablation-fast-slow}
\end{minipage}
\hfill
\begin{minipage}[t]{0.24\textwidth}
    \centering
    \includegraphics[width=\linewidth]{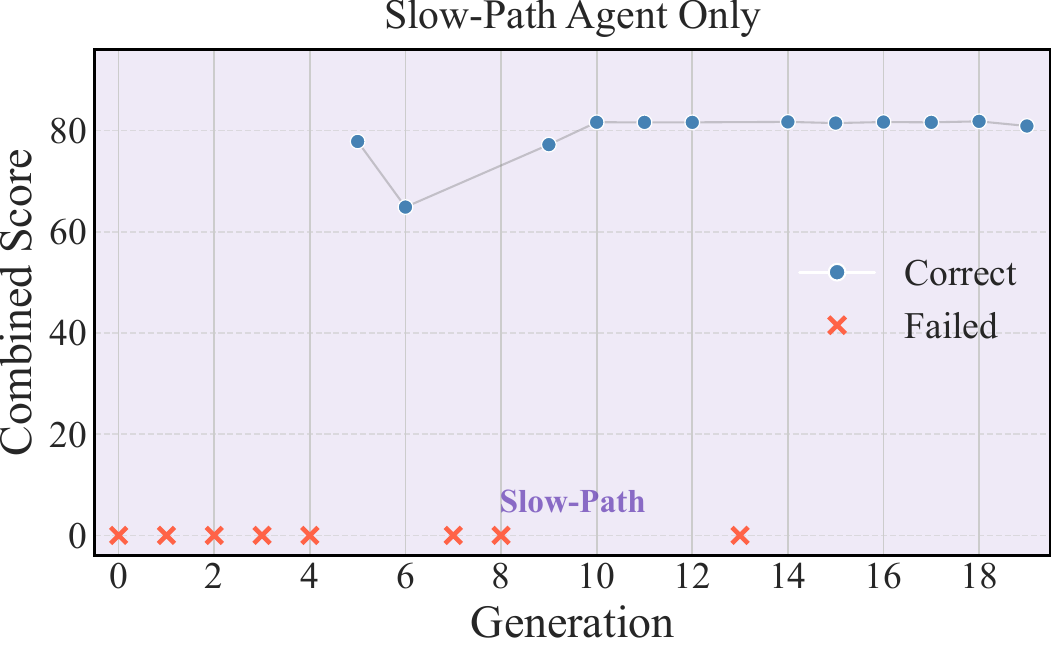}
    \captionof{figure}{\small{Slow-path agent only on the same workload.}}
    \label{fig:ablation-slow-only}
\end{minipage}
\hfill
\begin{minipage}[t]{0.24\textwidth}
    \centering
    \includegraphics[width=\linewidth]{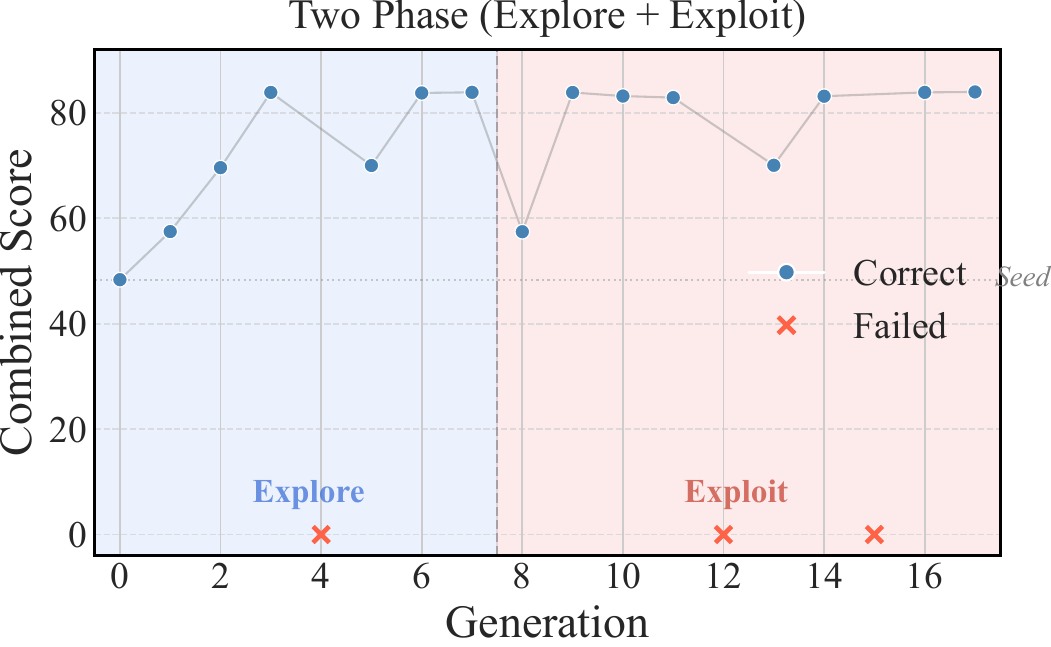}
    \captionof{figure}{\small{Two-phase evolution (explore-exploit) on the MoE workload.}}
    \label{fig:evo-two-phase}
\end{minipage}
\hfill
\begin{minipage}[t]{0.24\textwidth}
    \centering
    \includegraphics[width=\linewidth]{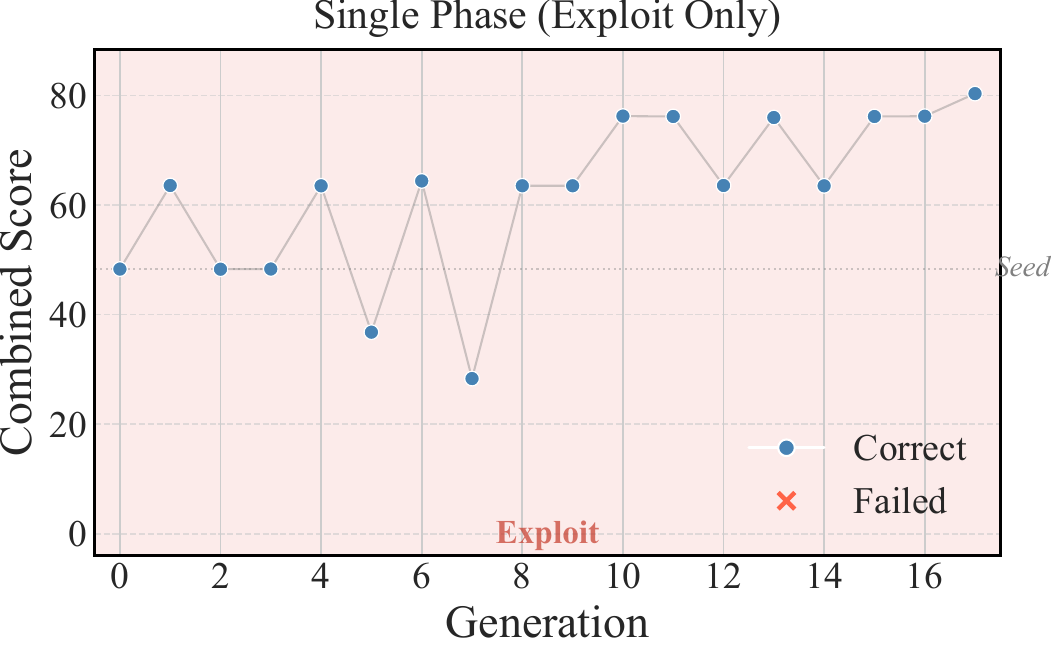}
    \captionof{figure}{\small{Single-phase evolution (exploit only) on the same workload.}}
    \label{fig:evo-single-phase}
\end{minipage}
\vspace{-12pt}
\end{figure*}

Figures~\ref{fig:ablation-fast-slow} and~\ref{fig:ablation-slow-only} show the results. With the fast-path agent, a correct GIN kernel emerges in 2~LLM iterations (score 48.34), and the slow-path reaches \textbf{83.95} within 3~generations. Without it (Figure~\ref{fig:ablation-slow-only}), the evolutionary agent must discover the GIN programming model while simultaneously optimizing---the first 5~generations all fail (compilation errors, deadlocks), wasting 25\% of the budget. It does not stabilize above 81 until generation~10 and peaks at \textbf{81.81}---\textbf{2.5\% lower}. The fast-path decomposition avoids wasting early generations on infrastructure discovery.

\noindent\textbf{Explore-Exploit Phases.}
We compare a \emph{two-phase} schedule (40\% explore then exploit) against \emph{single-phase} exploit-only, both on the GIN MoE workload with 18~generations. The two-phase schedule (Figure~\ref{fig:evo-two-phase}) reaches 83.86 by generation~3 and peaks at \textbf{83.95}. The exploit-only schedule (Figure~\ref{fig:evo-single-phase}) oscillates for 9~generations, exceeds 76 only at generation~10, and peaks at 80.36---a \textbf{4.3\% lower final score}. The explore phase discovers structurally diverse strategies (barrier-free overlap, split put/wait, signal-shadow synchronization) that the exploit phase refines; without this diversity, the agent converges slowly to a local optimum.

\noindent\textbf{Naive Iterative Prompting.}
We compare against iterative prompting without population, crossover, or meta-recommendations on the same GIN MoE workload (Figure~\ref{fig:ablation-naive}). Over 20~generations, naive prompting improves by only 1.1\% (144$\to$142.3\,ms), plateauing after generation~11. \cufuse{} reaches 118.1\,ms within 3~generations---a \textbf{17\% improvement}---confirming the value of population-based search over single-program refinement.

\vspace{-8pt}
\section{Conclusion}
\vspace{-8pt}
\cufuse{} automates compute-communication co-design by constraining an LLM agent to a structured configuration space rather than unconstrained code generation, separating correctness (fast-path transformation to a verified seed) from performance (evolutionary search), and injecting hardware context so the same pipeline adapts across topologies. Across four multi-GPU workloads, \cufuse{} delivers up to $1.57\times$ speedup over host-driven baselines and a 12--14\% per-layer advantage over standard DeepEP usage on intranode MoE. These results suggest that LLM agents are most effective as bounded variation operators over domain-defined spaces, and that correctness-first decomposition is essential to avoid wasting search budget on broken candidates.



%% file: appendix/main.tex
\appendix

\input{appendix/evolution}

%% file: appendix/evolution.tex
\section{NCCL Device-Initiated API Example}\label{sec:nccl_demo}

\begin{figure}[h]
  \centering
  \resizebox{0.7\linewidth}{!}{
\begin{lstlisting}[style=customc2]
template <typename T>
__global__ void HybridAlltoAllKernel(ncclWindow_t sendwin,
  size_t sendoffset, ncclWindow_t recvwin,
  size_t recvoffset, size_t count, int root,
  struct ncclDevComm devComm) {
  // 1. Initialize GIN context
  int ginContext = 0; // GIN context index
  unsigned int signalIndex = 0; // Signal slot
  ncclGin gin {devComm, ginContext};
  uint64_t signalValue = gin.readSignal(signalIndex);

  // 2. Cross-GPU synchronization before data exchange
  ncclBarrierSession<ncclCoopCta> bar {ncclCoopCta(),
    ncclTeamTagWorld(), gin, blockIdx.x};
  bar.sync(ncclCoopCta(), cuda::memory_order_relaxed,
    ncclGinFenceLevel::Relaxed);

  int tid = threadIdx.x + blockIdx.x * blockDim.x;
  int nthreads = blockDim.x * gridDim.x;

  // 3. Determine local (LSA) vs remote peers (GIN)
  ncclTeam world = ncclTeamWorld(devComm); // All ranks
  ncclTeam lsa = ncclTeamLsa(devComm); // Same-node ranks
  const int startLsa = world.rank - lsa.rank;
  const int lsaSize = lsa.nRanks;

  // 4. Send to remote peers via GIN (cross-node)
  const size_t size = count * sizeof(T);
  // Remote peers before LSA range
  for (int r=tid; r<startLsa; r+=nthreads) {
    gin.put(world, r, recvwin, recvoffset+world.rank*size,
      sendwin, sendoffset+r*size, size, ncclGin_SignalInc{signalIndex});
  }
  // Remote peers after LSA range
  for (int r=startLsa+lsaSize+tid; r<world.nRanks; r+=nthreads) {
    gin.put(world, r, recvwin,
      recvoffset+world.rank*size, sendwin, sendoffset+r*size,
      size, ncclGin_SignalInc{signalIndex});
  }

  // 5. Send to local peers via direct memory access (LSA)
  T* sendLocal = (T*)ncclGetLocalPointer(sendwin, sendoffset);
  for (size_t offset=tid; offset<count; offset+=nthreads) {
    for (int lp = 0; lp < lsa.nRanks; lp++) {
      int wr = startLsa + lp;
      T* recvPtr = (T*)ncclGetLsaPointer(recvwin, recvoffset, lp);
      recvPtr[world.rank*count+offset] = sendLocal[wr*count+offset];
    }
  }

  // 6. Wait for remote GIN operations to complete
  int numRemotePeers = world.nRanks - lsa.nRanks;
  gin.waitSignal(ncclCoopCta(), signalIndex, signalValue+numRemotePeers);
  gin.flush(ncclCoopCta());

  // 7. Final barrier to ensure all data is visible
  bar.sync(ncclCoopCta(), cuda::memory_order_release,
    ncclGinFenceLevel::Relaxed);
}
\end{lstlisting}}
  \caption{
    \small{\textbf{NCCL device-initiated API:} All-to-All CUDA kernel using the GPU-Initiated Networking (GIN) and Load/Store Accessible (LSA) API. GIN issues RDMA-style puts to remote peers (lines 31, 36); LSA accesses intra-node peer memory via direct loads and stores (lines 46--47). Fine-grained synchronization primitives (lines 12--15) and \texttt{gin.waitSignal} (line 53) enable compute--communication fusion that is out of reach for host-driven APIs.}
  }
  \label{fig:nccl_demo}
\end{figure}

\section{Design Space Details}\label{sec:design_space_details}

This section provides a full semantic breakdown of the eight configuration dimensions that constitute the optimization space $\mathcal{C} = \mathcal{B} \times \mathcal{M} \times \mathcal{P} \times \mathcal{S} \times \mathcal{I} \times \mathcal{G} \times \mathcal{O} \times \mathcal{K}$. The dimensions fall into two categories: \emph{concrete API dimensions} that map directly to fixed NCCL device-API identifiers and must be instantiated exactly, and \emph{intent-based dimensions} that express high-level optimization objectives the agent realizes freely during code generation.

\paragraph{Concrete API Dimensions.}

\begin{itemize}[leftmargin=1.4em,itemsep=2pt,topsep=2pt]

\item \textbf{Backend ($\mathcal{B}$): GIN $|$ LSA $|$ Hybrid.}
The backend determines the transport mechanism for device-initiated communication. \textbf{GIN} (GPU-Initiated Networking) provides one-sided RDMA put semantics: a thread group issues \texttt{ncclGinPut} to transfer data to a remote rank's registered window, then \texttt{ncclGinFlush} to guarantee delivery, and \texttt{ncclGinWaitSignal} to poll for a remote completion notification. GIN works across both intra-node (NVLink) and inter-node (InfiniBand/RoCE) links, making it the default choice for multi-node deployments. \textbf{LSA} (Load/Store Accessible) provides direct peer memory access within an NVLink domain: threads read from and write to remote buffers via standard load/store instructions resolved through \texttt{ncclGetLsaPointer}, synchronized by \texttt{ncclLsaBarrierSession}. LSA avoids network-stack overhead but is restricted to intra-node topologies. \textbf{Hybrid} combines both: LSA for intra-node transfers and GIN for inter-node transfers within the same kernel, requiring careful scope partitioning to avoid cross-backend interference.

\item \textbf{Completion Mechanism ($\mathcal{M}$): Barrier $|$ Signal $|$ SignalShadow $|$ Counter.}
The completion mechanism dictates how a rank detects that a remote transfer has finished. \textbf{Barrier} imposes a global rendezvous: all ranks in the synchronization scope must reach the barrier before any can proceed, guaranteeing collective completion at the cost of serializing all participants. \textbf{Signal} enables point-to-point notification: the sender increments the receiver's signal counter upon completion, and the receiver polls via \texttt{ncclGinWaitSignal}, allowing fine-grained overlap but introducing race-condition risks if signal ordering is not carefully managed. \textbf{SignalShadow} extends Signal by maintaining a shadow copy of the signal counter in local memory, reducing remote polling traffic at the cost of additional bookkeeping. \textbf{Counter} uses device-side atomic counters (\texttt{atomicAdd}/\texttt{atomicLoad}) for intra-kernel synchronization between communication and compute thread groups, enabling per-tile pipelining within a cooperative grid.

\item \textbf{Issuer Granularity ($\mathcal{I}$): Thread $|$ Warp $|$ WarpSpan $|$ Tile$\langle$N$\rangle$ $|$ CTA.}
The issuer granularity determines which thread group collectively participates in a single communication operation. \textbf{Thread}-level issuance offers maximum flexibility but incurs high per-operation overhead and risks warp divergence. \textbf{Warp}-level issuance amortizes setup across 32 threads and aligns with the GPU's native scheduling unit. \textbf{WarpSpan} allows a contiguous span of warps to issue cooperatively, useful for large transfers that exceed a single warp's bandwidth. \textbf{Tile$\langle$N$\rangle$} groups $N$ warps into a tile-sized unit matched to the compute tile dimensions, aligning communication granularity with compute structure. \textbf{CTA}-level issuance is the most conservative: the entire thread block participates, minimizing divergence risk but limiting overlap opportunities within the block.

\item \textbf{Memory Ordering ($\mathcal{O}$): Relaxed $|$ Acquire $|$ Release $|$ AcqRel.}
Memory ordering controls the visibility guarantees of device-initiated transfers. \textbf{Relaxed} ordering imposes no fence, allowing maximum reordering by the hardware for throughput but requiring the programmer (or agent) to ensure correctness through other means. \textbf{Release} ordering guarantees that all prior writes are visible before the transfer is signaled as complete---essential when a sender must ensure data integrity before notifying the receiver. \textbf{Acquire} ordering guarantees that subsequent reads observe all writes that preceded the matching release---essential on the receiver side after polling a signal. \textbf{AcqRel} combines both, providing a full fence suitable for bidirectional exchanges but at maximum ordering cost.

\item \textbf{Context Multiplicity ($\mathcal{K}$): 1 $|$ 2 $|$ N.}
Context multiplicity controls the number of concurrent communication contexts (NCCL device communicator instances) available to the kernel. A single context ($K{=}1$) serializes all transfers through one communicator, simplifying synchronization but limiting bandwidth utilization. Two contexts ($K{=}2$) enable double-buffering: one context transfers the current chunk while the other prepares the next, hiding transfer latency behind data preparation. $N$ contexts generalize this to deeper pipelines, useful for ring or mesh topologies where multiple concurrent peer transfers can saturate available network bandwidth.

\end{itemize}

\paragraph{Intent-Based Dimensions.}

\begin{itemize}[leftmargin=1.4em,itemsep=2pt,topsep=2pt]

\item \textbf{Communication Placement ($\mathcal{P}$): Deferred $|$ Tile-fused $|$ Tile-pipelined $|$ Stream-split.}
Placement determines where communication is positioned relative to compute. \textbf{Deferred} placement executes all communication after compute completes, the simplest and safest option but offering no overlap. \textbf{Tile-fused} interleaves communication with compute at the tile level within a single persistent kernel using warp specialization: dedicated communication warps issue transfers while compute warps process tiles, maximizing overlap for iterative exchange patterns (e.g., ring attention). \textbf{Tile-pipelined} issues communication for the next tile while compute processes the current tile, requiring double-buffering but enabling overlap without warp specialization. \textbf{Stream-split} separates communication onto a high-priority CUDA stream while compute runs on the default stream, relying on CUDA's multi-stream execution model for overlap; this is suited to bulk transfers between large compute phases where kernel-level fusion would be overly complex.

\item \textbf{Synchronization Scope ($\mathcal{S}$): Local $|$ World $|$ Rail $|$ Hierarchical.}
The synchronization scope defines the set of ranks that must participate in collective synchronization. \textbf{Local} scope restricts synchronization to ranks within the same node (NVLink domain), minimizing latency for intra-node operations. \textbf{World} scope includes all ranks across all nodes, necessary for global collectives but imposing cross-network synchronization costs. \textbf{Rail} scope synchronizes ranks that share the same GPU index across nodes (rail-optimized topology), exploiting the fact that NVSwitch fabrics provide uniform bandwidth within a rail. \textbf{Hierarchical} scope composes local and world synchronization in phases---first synchronizing within each node, then across nodes---reducing global synchronization frequency at the cost of additional implementation complexity.

\item \textbf{Transfer Granularity ($\mathcal{G}$): Per-peer $|$ Per-tile $|$ Per-chunk.}
Transfer granularity controls how data is partitioned for communication. \textbf{Per-peer} issues a single transfer per destination rank, minimizing operation count but preventing fine-grained overlap. \textbf{Per-tile} aligns transfer boundaries with compute tile boundaries, enabling per-tile pipelining where each tile can be processed as soon as its data arrives. \textbf{Per-chunk} subdivides each peer's data into smaller chunks, offering the finest-grained overlap at the cost of higher per-operation synchronization overhead.

\end{itemize}

\paragraph{Expert System Mapping.}
Table~\ref{tab:design_space_coverage} in the main text maps expert-crafted systems onto $\mathcal{C}$ by semantic equivalence. DeepEP's NVLink path uses direct peer memory stores with CTA-level barrier synchronization (LSA, Barrier, Deferred, Local, CTA, Per-peer, Release), while its InfiniBand path uses custom RDMA one-sided puts with signal-based completion (GIN, Signal, Deferred, World, CTA, Per-peer, Acquire, 1 context). FLUX employs persistent tile-fused kernels with warp-level LSA stores and bidirectional ordering (LSA, Barrier, Tile-fused, Local, Warp, Per-tile, AcqRel). These mappings demonstrate that the eight dimensions of $\mathcal{C}$ are sufficient to characterize the design decisions of state-of-the-art expert implementations, while exposing a much larger space of unexplored configurations that \cufuse{} can navigate automatically.

\section{Agent Context Details}\label{sec:agent_context}

The agent context is a structured prompt component assembled at runtime and injected into every agent invocation. It is \emph{conditioned on the selected backend} $\mathcal{B}$: a GIN run receives only GIN-specific documentation, headers, correctness rules, and reference code, while an LSA run receives the corresponding LSA-specific material. Shared components---the optimization strategy framework, evolve-block constraints, and hardware context---are included in both. This backend-conditioned assembly avoids polluting the agent's context window with irrelevant API surface, which we found reduced hallucinated cross-API calls in early experiments.

\paragraph{API Documentation.}
The context includes interface specifications following a fixed schema of function signatures, semantic preconditions, valid/invalid parameter combinations, and annotated usage examples. For a GIN run, this comprises the \emph{GIN interface} (one-sided RDMA put semantics, flush ordering, completion signaling), the \emph{team specification} (rank organization into world, local, and rail scopes, grounding $\mathcal{S}$), and the \emph{thread-group specification} (thread-level participation granularity, grounding $\mathcal{I}$). For an LSA run, the GIN interface is replaced by the \emph{LSA interface} (direct load/store access, memory barriers, intra-node peer connectivity). These establish the semantic contracts of the design space---details that cannot be inferred from general CUDA or host-driven NCCL experience alone. In addition, the context supplies the raw NCCL device-API C++ headers for the selected backend (e.g., \texttt{gin.h}, \texttt{core.h}, \texttt{coop.h} for GIN) and a complete, working reference kernel that demonstrates the target communication pattern end-to-end, giving the agent both a compilable type reference and a concrete pattern to generalize from rather than invent from scratch. Table~\ref{tab:lsa-example-appendix} shows the LSA interface specification as a representative example.

\begin{table}[h]
\centering
\scriptsize
\caption{Sample knowledge base structure as presented within the LSA interface.}
\label{tab:lsa-example-appendix}
\begin{tabular}{p{0.15\columnwidth} p{0.75\columnwidth}}
\toprule
\textbf{Field} & \textbf{Content} \\
\midrule
Purpose & Device-side peer memory access and barrier synchronization within local GPU teams via direct load/store. \\
Core API & \texttt{ncclLsaBarrierSession} (provides synchronization), \texttt{ncclGetLsaPointer} (team-indexed access), \texttt{ncclGetLocalPointer} (local access), ... \\
Sync Scope & \texttt{ncclTeamLsa} (intra-node local team), \texttt{ncclTeamWorld} (all ranks); tighter scope reduces overhead but limits overlap potential. \\
Invariants & Barrier index must be unique per CTA (\texttt{blockIdx.x}), ... \\
Canonical Pattern & Instantiate \texttt{ncclLsaBarrierSession} with \texttt{blockIdx.x}; perform memory operations; call \texttt{sync(ncclCoopCta(), memory\_order\_release)}; access peers via \texttt{ncclGetLsaPointer}. \\
\bottomrule
\end{tabular}
\end{table}

\paragraph{Strategy Knowledge.}
A shared optimization strategy framework spans both backends, guiding the agent to apply the right level of fusion for each workload. It describes three principal strategies:
\begin{itemize}[leftmargin=1.4em,itemsep=1pt,topsep=2pt]
\item \emph{Kernel-level fusion} --- persistent kernel with warp specialization or tile-and-send, suited to iterative fine-grained exchanges where per-round host overhead dominates.
\item \emph{Stream-level overlap} --- separate high-priority communication and compute streams, suited to bulk transfers between large compute phases where kernel fusion complexity is unwarranted.
\item \emph{Split put/wait} --- deferred completion with intervening compute, suited to pipelining where the sender can proceed with useful work after issuing the put but before confirming delivery.
\end{itemize}
The framework does not prescribe a preferred strategy; it articulates the conditions under which each approach dominates and leaves the architectural decision to the agent. Layered on top of this shared framework are \emph{backend-specific correctness rules}, conditioned on $\mathcal{B}$, that distill hard-won operational knowledge:
\begin{itemize}[leftmargin=1.4em,itemsep=1pt,topsep=2pt]
\item \textbf{GIN rules:} windows must use \texttt{ncclMemAlloc}; \texttt{waitSignal} polls this rank's own signal counter; \texttt{flush} must match the thread group that issued the puts; signal counters are monotonically increasing and must not be reset within a kernel.
\item \textbf{LSA rules:} barrier index should use \texttt{blockIdx.x} for CTA uniqueness; use \texttt{memory\_order\_release} after writes and \texttt{memory\_order\_acquire} before reads; all ranks in the LSA team must reach the barrier before any peer pointers are dereferenced.
\end{itemize}
These rules reduce the space of plausible-but-incorrect programs without constraining architectural exploration.

\paragraph{Hardware Context.}
Rather than hardcoding hardware descriptions, \cufuse{} dynamically extracts the hardware context from the evaluation harness's build and run configuration (compiler flags, MPI topology, NCCL version), mapping GPU architecture codes to known device properties via a lookup table. This makes the context portable across deployments: changing the target GPU or topology in the evaluation harness automatically updates the agent's hardware knowledge. The extracted context provides four categories:
\begin{itemize}[leftmargin=1.4em,itemsep=1pt,topsep=2pt]
\item \emph{GPU architecture}: device model, SM count, HBM bandwidth, shared-memory capacity, thread limits---grounding launch configuration and occupancy decisions.
\item \emph{Topology and interconnect}: number of ranks, inter- vs.\ intra-node placement, network type (InfiniBand, RoCE, NVLink), and peer-connectivity reachability---grounding backend selection $\mathcal{B}$ and synchronization scope $\mathcal{S}$.
\item \emph{Build environment}: CUDA and NCCL versions, compiler flags, MPI configuration---ensuring the agent generates code compatible with the toolchain.
\item \emph{Resource budgeting}: concrete SM-overhead estimates for communication kernels, NIC DMA independence properties for GIN, and guidance on how lightweight put/wait kernels are relative to the total SM budget---grounding placement $\mathcal{P}$ and issuer $\mathcal{I}$ decisions.
\end{itemize}
Together, these ensure that agent decisions---block counts, stream priorities, overlap strategy, backend choice---reflect the physical realities of the target deployment rather than training-time priors.

\section{Fast-Path Transformation Details}\label{sec:fast_path_details}

The fast-path agent converts a host-driven compute--communication program into a verified device-initiated implementation through a three-step pipeline. This section details each step.

\paragraph{CUDA Code Analysis.}
Because the input is user-written code rather than a formal specification, the agent must first recover a precise compute--communication boundary to ground the transformation. A lightweight static analyzer identifies all host-driven NCCL collectives, their buffer operands, data dependencies, and ordering constraints, producing a communication dependency graph. For each NCCL collective, the graph identifies: (1) the buffer operands and their allocation method (flagging \texttt{cudaMalloc} buffers that will need migration to \texttt{ncclMemAlloc} for device-initiated communication), (2) the producer and consumer kernels that define data-flow dependencies, and (3) the execution ordering constraints imposed by stream synchronization. This graph defines the transformation targets for later stages. Appendix~\ref{app:analysis_example} illustrates the analyzer's input and output on a representative MoE dispatch--compute--combine pipeline.

\paragraph{Host-to-Device Transformation.}
Using the dependency graph, the agent rewrites host-driven communication into device-initiated forms. The transformation proceeds in two stages, each guided by an LLM--judge feedback loop that ensures the rewritten program both compiles and matches the host baseline in correctness. The pipeline follows an empirical observation that converting host-driven NCCL to device-initiated communication naturally decomposes into two tasks: communication setup and semantic replacement. Solving both at once significantly increases the likelihood of correctness failures; separating them constrains each LLM interaction to a single concern, reducing the per-stage failure space and making judge feedback more targeted.

\noindent\underline{\emph{Stage~A: Communication Setup.}} The LLM constructs the host-side infrastructure required by the NCCL device API in three steps:
\begin{enumerate}[leftmargin=1.4em,itemsep=1pt,topsep=2pt]
\item \emph{Symmetric memory allocation}: All buffers identified by the analyzer as requiring device-initiated access are migrated from \texttt{cudaMalloc} to \texttt{ncclMemAlloc}, which allocates memory uniformly addressable across ranks.
\item \emph{Communicator configuration}: The communicator requirements are configured according to $\mathcal{B}$---using the LSA barrier count for $\mathcal{B}{=}\texttt{LSA}$, or the GIN barrier and signal counts for $\mathcal{B}{=}\texttt{GIN}$, reflecting each backend's distinct synchronization model (memory-based barriers vs.\ network-initiated signals).
\item \emph{Device communicator instantiation}: The device communicator is instantiated using the populated requirements and registered as a kernel argument, making the communication context available on-device.
\end{enumerate}
Compute and communication logic remain unchanged during Stage~A. The LLM judge verifies compilation and, on failure, identifies the root cause and feeds corrective feedback into the next rewrite iteration. Stage~A typically converges in 1--2 iterations.

\noindent\underline{\emph{Stage~B: Communication Replacement.}} With infrastructure in place, the LLM replaces each host-driven collective with its device-initiated equivalent. To prioritize correctness over performance, all directive dimensions are set conservatively:
\begin{itemize}[leftmargin=1.4em,itemsep=1pt,topsep=2pt]
\item CTA-level issuance ($\mathcal{I}{=}\texttt{CTA}$) to avoid warp-level divergence.
\item Barrier completion ($\mathcal{M}{=}\texttt{Barrier}$) to ensure deterministic collective synchronization.
\item World synchronization scope ($\mathcal{S}{=}\texttt{World}$) to ensure cross-rank visibility before compute resumes.
\item Deferred placement ($\mathcal{P}{=}\texttt{Deferred}$) to minimize ordering complexity.
\item Per-peer granularity ($\mathcal{G}{=}\texttt{Per\text{-}peer}$) to amortize per-operation overhead.
\item Release ordering ($\mathcal{O}{=}\texttt{Release}$) to guarantee write visibility.
\item Single context ($\mathcal{K}{=}1$) to avoid multi-context synchronization issues.
\end{itemize}
Backend-specific translation then maps host collectives to device primitives: for $\mathcal{B}{=}\texttt{GIN}$, host collectives map to \texttt{ncclGinPut} (one-sided network transfer), followed by \texttt{ncclGinWaitSignal} and \texttt{ncclGinFlush} to confirm remote delivery; for $\mathcal{B}{=}\texttt{LSA}$, to direct peer memory stores via \texttt{ncclGetLsaPointer} synchronized with \texttt{ncclLsaBarrierSession}. The LLM judge reviews each generated variant for compilation and correctness issues, summarizes the likely root causes, and feeds targeted guidance back to the model for a corrective regeneration. Using an LLM judge rather than rule-based error parsing is essential because CUDA failures often hinge on contextual reasoning---such as inferring which missing synchronization, ordering constraint, or API misuse triggered a particular error. Stage~B typically converges in 1--3 iterations.

\paragraph{Evolve-Block Annotation.}
The verified device-initiated program is automatically annotated with mutable-region markers that define the optimization scope for the slow-path agent. An annotator combining LLM analysis with pattern-based heuristics emits paired \emph{EVOLVE-BLOCK-START} / \emph{EVOLVE-BLOCK-END} markers around regions whose communication or compute structure admits optimization under the full constraint set $\mathcal{C}$. Frozen regions---initialization, verification logic, output formatting---are excluded, ensuring that downstream mutations cannot break the evaluation harness. Critically, the $n$ annotated regions are co-optimized as a single candidate rather than treated as independent subproblems, since mutations can interact across boundaries: a change to transfer granularity ($\mathcal{G}$) in one region may alter the synchronization requirements ($\mathcal{S}$) of another, or a shift from barrier to signal completion ($\mathcal{M}$) may require corresponding changes to memory ordering ($\mathcal{O}$) in a different region.

The resulting annotated program is promoted as the \emph{seed}---a correctness-preserving but intentionally unoptimized baseline that is passed to the slow-path agent as generation zero of the evolutionary search for performance improvements.

\section{Multi-Island Evolution Details}\label{sec:multi-island}

The slow-path agent uses a multi-island evolutionary framework to prevent premature convergence and maintain diversity across the search space $\mathcal{C}$.

\paragraph{Island Initialization.}
The search maintains $K$ independent islands, each initialized from a distinct seed---a semantically different variant of the fast-path baseline produced by prompting the LLM with different directive configurations drawn from $\mathcal{C}$. For example, one seed might use GIN with signal-based completion and stream-split placement, while another uses LSA with barrier completion and tile-fused placement. While a seed is a single starting program, an island is a full subpopulation of candidates evolved from that seed, maintaining its own evolutionary lineage. Each island evolves independently: one may converge on a multi-stream overlap strategy while another develops a fused cooperative-kernel approach entirely independently, ensuring that structurally different optimization strategies survive long enough to be evaluated on their merits.

\paragraph{Selection.}
Within each island, parent selection uses fitness-weighted sampling with configurable selection pressure, biasing reproduction toward high-performing candidates while preserving exploration breadth. The fitness score is the cascade evaluation score: $\text{score}(c) = 10000 / (1 + t_{\text{ms}})$ for candidates passing all three levels, or zero for candidates failing at $\ell_1$ or $\ell_2$. The selected parent, together with its stored evaluation history and LLM feedback, is passed to the search policy and mutation operator.

\paragraph{Migration.}
Every $G_m$ generations, the top-$k$ scoring individuals from each island are copied into randomly selected target islands, cross-pollinating successful patterns without collapsing diversity. Migration operates entirely in the program-text domain: high-scoring source code is directly inserted into the target island's population and subjected to the same LLM-guided mutation operators in subsequent generations. This design avoids the need for genotype encodings or crossover operators that would require structural alignment between candidates---the LLM naturally handles structural differences when it encounters a migrated program as a parent or archive inspiration.

\paragraph{MAP-Elites Archive.}
In addition to the per-island populations, the search maintains a global MAP-Elites diversity archive~\cite{mapelites}. The archive is a grid indexed by behavioral descriptors derived from the optimization directive (e.g., backend, placement strategy, completion mechanism), with each cell storing the highest-scoring candidate exhibiting that behavioral profile. Archive entries serve as cross-pollination inspirations during mutation: the LLM receives archive samples alongside the selected parent, exposing it to structurally distinct high-performing solutions from outside the current lineage. This mechanism is analogous to crossover across distant population members in traditional evolutionary algorithms, but operates at the semantic level of program structure rather than syntactic gene alignment.

\paragraph{Candidate Database.}
All evaluated candidates---including those eliminated at $\ell_1$ and $\ell_2$---are persisted to a shared candidate database. Each record stores the program text, the cascade score, the LLM feedback diagnostic, and a code embedding vector produced by a neural code encoder. The database serves two roles: (1) it is the candidate pool for parent selection and MAP-Elites archive sampling across all islands, and (2) it is the persistent backing store for the meta-summarizer (Appendix~\ref{sec:meta}), which queries historical results across all generations to distill cross-generation patterns into the mutation context.

Retrieval from the database is embedding-guided: at each generation, the current candidate's code embedding is used to query the database for the $k$ nearest neighbors by cosine similarity, surfacing structurally similar programs and their associated feedback. The retrieved records---both successes and failures---are injected into the LLM mutation prompt as in-context examples, allowing the agent to avoid previously failed transformations and build on patterns that improved performance in similar program structures.

\section{Static Analysis Example}
\label{app:analysis_example}

Listing~\ref{lst:host_moe} shows a simplified host-driven MoE pipeline following the DeepSeek-V3 dispatch--compute--combine pattern. Two \texttt{ncclAlltoAll} calls bracket an expert-compute phase: the first dispatches quantized tokens to remote experts, the second returns the processed results. All communication is host-driven and strictly sequential---no overlap with compute is possible.

\begin{lstlisting}[style=customc2,caption={Simplified host-driven MoE pipeline (input to the static analyzer).},label={lst:host_moe}]
void run_moe(int rank, int nranks, ncclComm_t comm) {
  cudaStream_t stream;
  cudaStreamCreate(&stream);

  // --- Buffers (cudaMalloc, not device-comm compatible) ---
  int8_t *d_quant_send, *d_quant_recv;
  float  *d_expert_out, *d_final_out;
  cudaMalloc(&d_quant_send, chunk_bytes);
  cudaMalloc(&d_quant_recv, chunk_bytes);
  cudaMalloc(&d_expert_out, out_bytes);
  cudaMalloc(&d_final_out,  out_bytes);

  // --- Quantize tokens ---
  quantize<<<grid, block, 0, stream>>>(
      d_input, d_quant_send, d_scales, num_tokens, HIDDEN);

  // --- Dispatch: host-driven AlltoAll (int8) ---
  ncclAlltoAll(d_quant_send, d_quant_recv,
               chunk_elems, ncclInt8, comm, stream);
  cudaStreamSynchronize(stream);

  // --- Expert compute ---
  dequantize<<<grid, block, 0, stream>>>(
      d_quant_recv, d_deq, d_scales, num_tokens, HIDDEN);
  gemm<<<grid, block, 0, stream>>>(
      d_deq, d_W1, d_gemm1, tokens, GEMM1_DIM, HIDDEN);
  swiGLU<<<grid, block, 0, stream>>>(
      d_gemm1, d_swiglu, tokens, INTER_DIM);
  gemm<<<grid, block, 0, stream>>>(
      d_swiglu, d_W2, d_expert_out, tokens, HIDDEN, INTER_DIM);

  // --- Combine: host-driven AlltoAll (float) ---
  ncclAlltoAll(d_expert_out, d_final_out,
               chunk_elems, ncclFloat, comm, stream);
  cudaStreamSynchronize(stream);
}
\end{lstlisting}

The static analyzer processes Listing~\ref{lst:host_moe} and produces the communication dependency graph shown in Listing~\ref{lst:comm_graph}. For each NCCL collective, the graph identifies the buffer operands, their allocation method (flagging \texttt{cudaMalloc} buffers that will need migration to \texttt{ncclMemAlloc} for device-initiated communication), and the producer/consumer kernels---establishing the data-flow context that subsequent pipeline steps use to determine transformation targets.

\begin{lstlisting}[style=customc2,caption={Communication dependency graph produced by the static analyzer for Listing~\ref{lst:host_moe}.},label={lst:comm_graph}]
Communication Graph
  Node 1: ncclAlltoAll (line 19)
    Stream: stream
    Count: chunk_elems, Datatype: ncclInt8

    Send buffer: d_quant_send
      Allocated: cudaMalloc (line 8) [needs ncclMemAlloc]
      Produced by: quantize<<<...>>> (line 15)
    Recv buffer: d_quant_recv
      Allocated: cudaMalloc (line 9) [needs ncclMemAlloc]
      Consumed by: dequantize<<<...>>> (line 24)

    Device-side transformation:
      Pattern: AlltoAll -- rank r sends sendbuf[peer*chunk]
               to peer's recvbuf[r*chunk] + local self-copy
      Buffers needing ncclMemAlloc: d_quant_send, d_quant_recv

  Node 2: ncclAlltoAll (line 34)
    Stream: stream
    Count: chunk_elems, Datatype: ncclFloat

    Send buffer: d_expert_out
      Allocated: cudaMalloc (line 10) [needs ncclMemAlloc]
      Produced by: gemm<<<...>>> (line 30)
    Recv buffer: d_final_out
      Allocated: cudaMalloc (line 11) [needs ncclMemAlloc]

    Device-side transformation:
      Pattern: AlltoAll -- rank r sends sendbuf[peer*chunk]
               to peer's recvbuf[r*chunk] + local self-copy
      Buffers needing ncclMemAlloc: d_expert_out, d_final_out

Execution Order
  run_moe: quantize<<<...>>> [compute]
           -> ncclAlltoAll [communicate]
           -> dequantize<<<...>>> [compute]
           -> gemm<<<...>>> [compute]
           -> swiGLU<<<...>>> [compute]
           -> gemm<<<...>>> [compute]
           -> ncclAlltoAll [communicate]
\end{lstlisting}

\section{Optimization Directive Listing}\label{sec:directive_listing}

The following shows the literal directive syntax that each agent emits before generating kernel code. Each field selects one value from the corresponding dimension of the configuration space $\mathcal{C}$ (Table~\ref{tab:fusion_space} in the main text).

\begin{lstlisting}[style=customc2]
optimization_directive:
  backend:    <GIN | LSA | Hybrid>
  completion: <Barrier | Signal | SignalShadow | Counter>
  sync_scope: <Local | World | Rail | Hierarchical>
  issuer:     <Thread | Warp | WarpSpan | Tile<N> | CTA>
  ordering:   <Relaxed | Acquire | Release | AcqRel>
  contexts:   <1 | 2 | 4>
  placement:  <intent: overlap aggressiveness>
  chunk_size: <intent: transfer granularity>
\end{lstlisting}

\section{Slow-Path Evolutionary Search Algorithm}\label{sec:slowpath_algo}

Algorithm~\ref{alg:slowpath} provides the formal pseudocode for the slow-path evolutionary search described in Section~\ref{subsec:slow_path}. The outer loop iterates over generations; within each generation, every island selects a parent, applies the LLM mutation operator (Algorithm~\ref{alg:llmmutate_appendix}), evaluates the offspring through the cascade, and updates the population, archive, and database. Periodically, high-scoring candidates migrate between islands.

\begin{algorithm}[H]
\caption{\cufuse{} Slow-Path: LLM-Driven Evolutionary Search}
\label{alg:slowpath}
\begin{algorithmic}[1]
\setlength{\itemsep}{0pt}
\setlength{\parskip}{0pt}
\Require{Baseline kernel $c_0 \in \mathcal{C}$, island count $k$, generation budget $G$, explore fraction $\alpha \in (0,1)$}
\Ensure{Optimized kernel $c^* \in \mathcal{C}$}
\State $\mathcal{P}_i \gets \textsc{Init}(c_0)$ for $i = 1, \ldots, k$
\State $\mathcal{DB},\ \mathcal{A} \gets \emptyset, \emptyset$
\For{$g \gets 1$ \textbf{to} $G$}
    \For{$i \gets 1$ \textbf{to} $k$}
        \State $p \gets \textsc{Select}(\mathcal{P}_i)$
        \State $c' \gets \textsc{LLMMutate}(p,\ \mathcal{A},\ \mathcal{DB},\ g,\ G,\ \alpha)$ \Comment{Alg.~\ref{alg:llmmutate_appendix}}
        \State $s \gets \textsc{CascadeEval}(c')$ \Comment{$\ell_1$: compile $\to$ $\ell_2$: verify $\to$ $\ell_3$: bench}
        \State $\mathcal{P}_i,\ \mathcal{A},\ \mathcal{DB} \gets \textsc{Update}(\mathcal{P}_i,\ \mathcal{A},\ \mathcal{DB},\ c',\ s)$
    \EndFor
    \State $\textsc{Migrate}(\mathcal{P}_1, \ldots, \mathcal{P}_k)$
\EndFor
\State \Return $c^* \gets \arg\max_{c\ \in\ \mathcal{DB}}\ s(c)$
\end{algorithmic}
\end{algorithm}

\section{LLMMutate: Phase-Dependent Variation Operator}\label{sec:llmmutate_details}

Algorithm~\ref{alg:llmmutate_appendix} details the \textsc{LLMMutate} operator used by the slow-path evolutionary search (Algorithm~\ref{alg:slowpath}). The operator selects an explore or exploit phase based on the current generation relative to the budget, constructs a context from the parent, archive, and meta-summarizer, and dispatches to one of three mutation forms at a phase-appropriate temperature.

\begin{algorithm}[H]
\caption{\textsc{LLMMutate}: Phase-Dependent Variation Operator}
\label{alg:llmmutate_appendix}
\begin{algorithmic}[1]
\setlength{\itemsep}{0pt}
\setlength{\parskip}{0pt}
\Require{Parent $p \in \mathcal{C}$, archive $\mathcal{A}$, database $\mathcal{DB}$, generation $g$, budget $G$, explore fraction $\alpha$}
\Ensure{Offspring $c' \in \mathcal{C}$}
\State $\phi \gets \begin{cases} \textsc{Explore} & g \leq \alpha G \\ \textsc{Exploit} & g > \alpha G \end{cases}$
\State $\text{ctx} \gets (p,\ \textsc{ArchiveSample}(\mathcal{A}),\ \textsc{MetaSummarize}(\mathcal{DB}))$
\If{$\phi = \textsc{Explore}$}
    \State $\text{form} \sim \text{Categorical}(\textsc{Rewrite} \gg \textsc{Diff},\ \textsc{Crossover})$
    \State $c' \gets \textsc{LLM}(\text{ctx},\ \text{form},\ \tau_{\text{high}})$
\Else
    \State $\text{form} \sim \text{Categorical}(\textsc{Diff} \gg \textsc{Rewrite},\ \textsc{Crossover})$
    \State $c' \gets \textsc{LLM}(\text{ctx},\ \text{form},\ \tau_{\text{low}})$
\EndIf
\State \Return $c'$
\end{algorithmic}
\end{algorithm}

\section{Novelty Filtering}
\noindent\textbf{Novelty Filtering.}
To prevent population collapse to near-identical programs, the search engine applies embedding-based novelty filtering. Each candidate's code is embedded via a neural code encoder, and candidates whose cosine similarity to any existing database entry exceeds a configurable threshold are rejected and resampled. This mechanism, combined with the multi-island structure and MAP-Elites archive, maintains structural diversity throughout the search and prevents the optimizer from repeatedly proposing minor lexical variants of the current best.

\section{Meta-Summarizer}\label{sec:meta}

The meta-summarizer is a periodic LLM-driven analysis module that extracts optimization patterns from accumulated evolution history and injects them back into the search as actionable recommendations. Every $k$ generations, it executes a three-step pipeline: (i)~\emph{summarize} the most recent batch of candidates---their scores, mutation types, architectural choices, and evaluation feedback---into a concise digest; (ii)~\emph{update} a persistent global scratchpad that tracks which strategies have been attempted, which succeeded, and which failed across the full evolution history; and (iii)~\emph{recommend} a ranked set of concrete optimization directions for the next generation, grounded in the accumulated evidence.

These recommendations are injected into the system prompt consumed by the LLM mutation operator, providing generation-over-generation learning without modifying the optimizer's architecture or search parameters. The meta-summarizer enables the search to develop an evolving understanding of what works for a specific workload and hardware target: early recommendations may suggest exploring different fusion levels, while later recommendations---informed by observing that multi-stream overlap consistently outperforms full fusion for a particular communication pattern---may redirect effort toward refining that specific strategy. This closed-loop meta-learning distinguishes the slow-path agent from static evolutionary code search: the mutation operator's effective behavior adapts over the course of evolution based on empirical evidence rather than fixed heuristics.

\section{Search Cost Details}\label{sec:search-cost-details}

Table~\ref{tab:search-cost} provides a per-workload breakdown of the search cost, including fast-path iterations, slow-path generations, wall-clock time, and total LLM API fees.

\begin{table}[h]
\centering
\scriptsize
\setlength{\tabcolsep}{5pt}
\caption{Search cost per workload. All runs use Claude Sonnet~4.5.}
\label{tab:search-cost}
\begin{tabular}{lcccc}
\toprule
\textbf{Workload} & \textbf{Fast-Path Iter.} & \textbf{Slow-Path Gen.} & \textbf{Wall Time} & \textbf{LLM Cost} \\
\midrule
Flash Attention   & 3 & 20 & ${\sim}$1.8\,hr & \$10.96 \\
MoE (GIN)        & 2 & 18 & ${\sim}$1.5\,hr & \$10.20 \\
KV-Cache         & 2 & 18 & ${\sim}$1.5\,hr & \$9.74 \\
GEMM + AllGather & 2 & 10 & ${\sim}$51\,min & \$4.98 \\
\bottomrule
\end{tabular}
\end{table}

\section{Workload Details}\label{sec:workload_details}

This section provides full descriptions of the four evaluation workloads, including host baseline implementations and the co-design pipelines discovered by \cufuse{}.

\paragraph{Flash Attention with Context Parallelism.}
Flash Attention with Context Parallelism scales attention over long sequences by partitioning the query, key, and value tensors along the sequence dimension across multiple GPUs. Each GPU permanently owns one Q shard while KV shards rotate in a ring topology (Ring Attention~\cite{liu2024ringattention}). As each GPU computes attention over its local KV tile, it concurrently receives the next KV block from its neighbor. Inspired by GPT-2's multi-head attention architecture, we evaluate on 4 GPUs (intra-node, NVLink) with configurations spanning $\text{SEQUENCE\_LENGTH} \in \{4096, 8192\}$ and $\text{HEAD\_DIM} \in \{32, 64\}$.

\paragraph{DeepSeek-V3 MoE Dispatch and Combine.}
We reproduce the MoE expert parallelism layer of DeepSeek-V3, in which each GPU routes its tokens to remote experts via an AlltoAll dispatch, runs expert computation, and gathers results via a reverse AlltoAll combine. The pipeline consists of: (1)~quantize tokens to int8, (2)~dispatch via AlltoAll, (3)~dequantize + GEMM1 (up/gate projection, $7168 \times 4096$) + SwiGLU activation + GEMM2 (down projection, $2048 \times 7168$) on the expert GPU, and (4)~combine results via reverse AlltoAll. The host baseline uses two-stream overlap with padded \texttt{ncclAlltoAll} on a dedicated communication stream. The \cufuse{} version replaces this with GIN one-sided RDMA puts using split put/wait kernels on separate streams, enabling variable-size per-peer transfers that avoid padding overhead, and overlapping dispatch with self-chunk expert compute concurrently with the network transfer. We evaluate on 2~GPUs across inter-node RoCE links with 4096~total tokens per rank under four sparse routing imbalance ratios (2:1, 3:1, 4:1, 5:1) to characterize the sensitivity of GIN to workload skew.

\paragraph{KV-Cache Transfer.}
KV-cache transfer in disaggregated prefill--decode serving requires the prefill GPU to compute and send the K and V projections before the decode stage can run attention. The flow is: compute K, send it, compute V while K is in flight, then send V with a readiness signal. The main bottleneck is the compute-to-send gap introduced by NCCL's CPU-launched transfers, which leave the network idle after each compute phase. The \cufuse{} path removes this gap by chaining a K GEMM, a GPU-triggered send, an overlapping V GEMM, and a final send with a signal increment; the decode GPU waits entirely on-device. We evaluate on two NVLink-connected GPUs with hidden dimension 4096.

\paragraph{GEMM + AllGather.}
Each rank performs a local FP32 GEMM ($C = AB$, $M{=}N{=}K{=}4096$) and then participates in an AllGather so every GPU receives the full concatenated output. This workload isolates the simplest post-compute collective: once a rank completes its GEMM, its output can be broadcast immediately with no cross-rank dependencies, enabling clear opportunities for device-initiated communication. Intra-node transfers over NVLink use LSA, while inter-node transfers over RoCE use GIN. We evaluate on four GPUs spanning both intra- and inter-node links.

\section{Flash Attention Case Study: Full Architecture}\label{sec:flash_attn_details}

This section provides the full architectural details of the \cufuse{}-evolved Flash Attention kernel and the host baseline, complementing the summary in Section~\ref{sec:flash-attn-deep-dive}.

\paragraph{GIN Fused Kernel Architecture.}
The \cufuse{} evolved kernel launches a single cooperative grid of $B \times n_h + 1 = 81$ blocks via \texttt{cudaLaunchCooperativeKernel}. Block~0 is the \emph{communication block}: it runs the entire ring protocol---GIN put/wait/flush for all 256 tiles across all 3 exchange rounds---using device-side atomic counters for synchronization. The remaining 80 blocks are \emph{compute blocks}, each handling one (batch, head) pair across all rounds. KV data is stored in a tile-first memory layout and double-buffered between two NCCL-registered windows.

The critical capability is \emph{per-tile pipelining}: as the comm block completes the transfer of tile~$j$ and increments a \texttt{tile\_ready} counter via \texttt{atomicAdd}, compute blocks immediately begin processing tile~$j$ by polling with \texttt{\_\_nanosleep(100)}---even while tiles $j\!+\!1$ through $T_c$ are still in flight. This eliminates the pipeline bubble between communication and computation within each round.

\paragraph{Host Baseline Architecture.}
The host version launches a separate attention kernel per round and issues per-tile \texttt{ncclGroup} calls from the CPU between rounds---768 host NCCL API call groups in total. Because the attention kernel (80 blocks) saturates all SMs, a small signaling kernel on the communication stream can never be scheduled---creating a GPU-side priority inversion that prevents per-tile pipelining. The host version must therefore wait for \emph{all} tiles to finish exchanging before launching the next round's attention kernel, forcing a purely sequential execution pattern.

%% file: checklist.tex
\section*{NeurIPS Paper Checklist}

\begin{enumerate}

\item {\bf Claims}
    \item[] Question: Do the main claims made in the abstract and introduction accurately reflect the paper's contributions and scope?
    \item[] Answer: \answerYes{}
    \item[] Justification: The abstract and introduction (Section~1) state three claims: (i) up to $1.57\times$ speedup over host-driven baselines, (ii) 12--14\% per-layer advantage over sequential DeepEP usage on MoE, and (iii) LLM cost under \$10 per workload. All three are supported by experimental results in Section~4. The introduction also explicitly states the target regime where co-design helps and where it does not.

\item {\bf Limitations}
    \item[] Question: Does the paper discuss the limitations of the work performed by the authors?
    \item[] Answer: \answerYes{}
    \item[] Justification: Section~1 explicitly identifies the target regime and states where co-design is less likely to help (compute-dominated workloads, negligible intra-node communication, or when an expert library already provides the right fused schedule). Section~4.3 discusses limitations of the DeepEP comparison, noting that in communication-dominated regimes DeepEP's faster raw transport would reduce the overlap window. The evaluation is conducted on four workloads on a specific hardware configuration (A100, NVLink/RoCE).

\item {\bf Theory assumptions and proofs}
    \item[] Question: For each theoretical result, does the paper provide the full set of assumptions and a complete (and correct) proof?
    \item[] Answer: \answerNA{}
    \item[] Justification: The paper does not include theoretical results or proofs. It is a systems paper presenting an empirical framework and experimental evaluation.

    \item {\bf Experimental result reproducibility}
    \item[] Question: Does the paper fully disclose all the information needed to reproduce the main experimental results of the paper to the extent that it affects the main claims and/or conclusions of the paper (regardless of whether the code and data are provided or not)?
    \item[] Answer: \answerYes{}
    \item[] Justification: Section~4 specifies the hardware (A100 80GB, NVLink, RoCE), software (CUDA 13.1, NCCL 2.28.9, Ubuntu 22.04.4), LLM model (Claude Sonnet 4.5), and all workload parameters (sequence lengths, head dimensions, token counts, skew ratios). The appendix provides full algorithmic pseudocode (Algorithms~1--2), design-space details, agent context structure, and workload descriptions sufficient to reimplement the system.

\item {\bf Open access to data and code}
    \item[] Question: Does the paper provide open access to the data and code, with sufficient instructions to faithfully reproduce the main experimental results, as described in supplemental material?
    \item[] Answer: \answerYes{}
    \item[] Justification: The authors will fully open-source \cufuse{}, including the framework code, workload implementations, and evaluation harness.

\item {\bf Experimental setting/details}
    \item[] Question: Does the paper specify all the training and test details (e.g., data splits, hyperparameters, how they were chosen, type of optimizer) necessary to understand the results?
    \item[] Answer: \answerYes{}
    \item[] Justification: Section~4 specifies the full environment setup (hardware, OS, CUDA/NCCL versions, LLM model). Evolutionary hyperparameters (population size, island count, explore fraction, migration interval, selection pressure) are detailed in the appendix (Appendix~\ref{sec:multi-island}). Workload parameters (matrix sizes, token counts, sequence lengths, head dimensions) are specified per workload in Section~4 and Appendix~\ref{sec:workload_details}.

\item {\bf Experiment statistical significance}
    \item[] Question: Does the paper report error bars suitably and correctly defined or other appropriate information about the statistical significance of the experiments?
    \item[] Answer: \answerNo{}
    \item[] Justification: We report median latency over five stable runs measured with CUDA events (Section~4.1), which is standard practice for GPU kernel benchmarking where run-to-run variance is negligible (sub-1\% for deterministic kernels on dedicated hardware). We do not report error bars because the variance across runs is smaller than the marker size in our plots. The evolutionary search is inherently stochastic, but we report the best-found solution rather than characterizing the distribution of search outcomes.

\item {\bf Experiments compute resources}
    \item[] Question: For each experiment, does the paper provide sufficient information on the computer resources (type of compute workers, memory, time of execution) needed to reproduce the experiments?
    \item[] Answer: \answerYes{}
    \item[] Justification: Section~4 details the hardware (two servers, each with 4$\times$ A100 80GB GPUs, dual Intel Xeon Silver 4314 CPUs, 256 GiB RAM, RoCE inter-node). Table~\ref{tab:search-cost} in the appendix reports per-workload wall time and LLM API cost. Each evolution run takes 45 minutes to 2 hours and costs \$5--\$11 in API fees.

\item {\bf Code of ethics}
    \item[] Question: Does the research conducted in the paper conform, in every respect, with the NeurIPS Code of Ethics \url{https://neurips.cc/public/EthicsGuidelines}?
    \item[] Answer: \answerYes{}
    \item[] Justification: This work optimizes GPU kernel performance for distributed ML workloads. It does not involve human subjects, personal data, or deceptive practices. The research conforms with the NeurIPS Code of Ethics.

\item {\bf Broader impacts}
    \item[] Question: Does the paper discuss both potential positive societal impacts and negative societal impacts of the work performed?
    \item[] Answer: \answerNA{}
    \item[] Justification: This work is foundational systems research on GPU kernel optimization. It improves computational efficiency of distributed ML training and inference, reducing energy consumption and hardware requirements. We do not foresee direct negative societal impacts beyond those generally associated with making ML training more efficient (which could accelerate both beneficial and harmful applications of ML). The technology does not introduce new capabilities for harm beyond what already exists.

\item {\bf Safeguards}
    \item[] Question: Does the paper describe safeguards that have been put in place for responsible release of data or models that have a high risk for misuse (e.g., pre-trained language models, image generators, or scraped datasets)?
    \item[] Answer: \answerNA{}
    \item[] Justification: The paper does not release pre-trained models, datasets, or other artifacts with high misuse risk. The released code is a kernel optimization framework that generates CUDA kernels for specific hardware configurations and poses no dual-use risk.

\item {\bf Licenses for existing assets}
    \item[] Question: Are the creators or original owners of assets (e.g., code, data, models), used in the paper, properly credited and are the license and terms of use explicitly mentioned and properly respected?
    \item[] Answer: \answerYes{}
    \item[] Justification: We cite all external libraries and systems used (NCCL, DeepEP, FLUX, NVSHMEM) with their original papers. We use the Claude Sonnet 4.5 API through Anthropic's commercial API under standard terms of service. NCCL is used under its BSD license.

\item {\bf New assets}
    \item[] Question: Are new assets introduced in the paper well documented and is the documentation provided alongside the assets?
    \item[] Answer: \answerYes{}
    \item[] Justification: We will release \cufuse{} as open-source software with documentation covering installation, workload definitions, and usage instructions. The four evaluation workloads and their host baselines will be included as reproducible examples.

\item {\bf Crowdsourcing and research with human subjects}
    \item[] Question: For crowdsourcing experiments and research with human subjects, does the paper include the full text of instructions given to participants and screenshots, if applicable, as well as details about compensation (if any)?
    \item[] Answer: \answerNA{}
    \item[] Justification: This paper does not involve crowdsourcing or research with human subjects.

\item {\bf Institutional review board (IRB) approvals or equivalent for research with human subjects}
    \item[] Question: Does the paper describe potential risks incurred by study participants, whether such risks were disclosed to the subjects, and whether Institutional Review Board (IRB) approvals (or an equivalent approval/review based on the requirements of your country or institution) were obtained?
    \item[] Answer: \answerNA{}
    \item[] Justification: This paper does not involve human subjects research.

\item {\bf Declaration of LLM usage}
    \item[] Question: Does the paper describe the usage of LLMs if it is an important, original, or non-standard component of the core methods in this research? Note that if the LLM is used only for writing, editing, or formatting purposes and does \emph{not} impact the core methodology, scientific rigor, or originality of the research, declaration is not required.
    \item[] Answer: \answerYes{}
    \item[] Justification: LLMs are the core component of \cufuse{}'s methodology. The paper clearly states that all agents (fast-path, slow-path, and feedback) use Anthropic's Claude Sonnet 4.5 (Section~4). The paper describes in detail how the LLM is used: as a mutation operator in evolutionary search (Section~3.3), as a code transformation agent (Section~3.2), and as a meta-summarizer (Appendix~\ref{sec:meta}). The LLM is not fine-tuned; it is used via API inference only.

\end{enumerate}